\documentclass{SciPost}

\hypersetup{
    colorlinks,
    linkcolor={red!50!black},
    citecolor={blue!50!black},
    urlcolor={blue!80!black}
}

\usepackage[bitstream-charter]{mathdesign}
\DeclareSymbolFont{usualmathcal}{OMS}{cmsy}{m}{n}
\DeclareSymbolFontAlphabet{\mathcal}{usualmathcal}

\fancypagestyle{SPstyle}{
\fancyhf{}
\lhead{\colorbox{scipostblue}{\bf \color{white} ~SciPost Physics }}
\rhead{{\bf \color{scipostdeepblue} ~Submission }}

\fancyfoot[C]{\textbf{\thepage}}
}

\DeclareMathOperator{\E}{E}
\DeclareMathOperator{\sgn}{sgn}
\newcommand{\dd}{\text{d}}
\newcommand{\ee}{\text{e}}

\usepackage{multirow}
\usepackage{dcolumn}
\newcolumntype{.}{D{.}{.}{2.8}}
\newcolumntype{(}{D{(}{(}{-1}}

\begin{document}

\pagestyle{SPstyle}

\begin{center}{\Large \textbf{\color{scipostdeepblue}{
On the true low-energy excitations of\\ the three-dimensional spin glass
}}}\end{center}

\begin{center}\textbf{
Claudio Chilin\textsuperscript{1,2 $\star$},
Enzo Marinari\textsuperscript{2,3},
Víctor Martín-Mayor\textsuperscript{1},
Giorgio Parisi\textsuperscript{4,2,3},
Juan~J.~Ruiz-Lorenzo\textsuperscript{5,6} 
and
David Yllanes\textsuperscript{7,8,9}
}\end{center}

\begin{center}
{\bf 1} Departamento de Física Teórica, Universidad Complutense, 28040 Madrid, Spain
{\bf 2} Dipartimento di Fisica, Sapienza Università di Roma, 00185 Rome, Italy
{\bf 3} CNR-Nanotec, Rome Unit, and INFN, Sezione di Roma, 00185 Rome, Italy
{\bf 4} International Research Center of Complexity Sciences, Hangzhou International Innovation Institute, Beihang University, Hangzhou 311115, China
{\bf 5} Departamento de Física, Universidad de Extremadura, 06006 Badajoz, Spain
{\bf 6} Instituto de Computación Científica Avanzada (ICCAEx), Universidad de Extremadura, 06006 Badajoz, Spain
{\bf 7} Fundación ARAID, Diputación General de Aragón, 50018 Zaragoza, Spain
{\bf 8} Instituto de Biocomputación y Física de Sistemas Complejos (BIFI), Universidad de Zaragoza, 50018 Zaragoza, Spain
{\bf 9} Zaragoza Scientific Center for Advanced Modeling (ZCAM), 50018 Zaragoza, Spain
\\[\baselineskip]
$\star$ \href{mailto:email1}{\small clachili@ucm.es}
\end{center}

\section*{\color{scipostdeepblue}{Abstract}}
\textbf{\boldmath{%
We study the low-energy excitations of the three dimensional spin glass
through a large-scale Monte Carlo
simulation on lattices up to $L=18$.   We find  smooth extrapolations down to zero temperature,  which, in the case of the energy and of the link overlap, can be directly ---and favourably--- compared with previous investigations featuring ground states (\emph{i.e.}, at zero temperature).  
The best fit for the fractal dimension of the excitations is
provided by Replica-Symmetry Breaking theory, but we also consider the alternative TNT description. The $P(q)$ is found to verify the Parisi-Toulouse temperature scaling.  Our data provides a spectacular confirmation of the overlap-equivalence hypothesis.
}}

\vspace{\baselineskip}

\noindent\textcolor{white!90!black}{%
\fbox{\parbox{0.975\linewidth}{%
\textcolor{white!40!black}{\begin{tabular}{lr}%
  \begin{minipage}{0.6\textwidth}%
    {\small Copyright attribution to authors. \newline
    This work is a submission to SciPost Physics. \newline
    License information to appear upon publication. \newline
    Publication information to appear upon publication.}
  \end{minipage} & \begin{minipage}{0.4\textwidth}
    {\small Received Date \newline Accepted Date \newline Published Date}%
  \end{minipage}
\end{tabular}}
}}
}


\vspace{10pt}
\noindent\rule{\textwidth}{1pt}
\tableofcontents
\noindent\rule{\textwidth}{1pt}
\vspace{10pt}

\newcommand{\tavg}[1]{\langle {#1} \rangle}
\newcommand{\davg}[1]{\overline {\tavg{#1}} }
\newcommand{\SDY}{\mathcal{S}}
\newcommand{\ql}{\ensuremath{q_\text{l}}}
\newcommand{\df}{\ensuremath{d_\text{f}}}
\newcommand{\qlqsq}{\E(\ql|q^2)}
\newcommand{\qlqz}{\E(\ql|q^2=0)}

\section{Introduction}
\label{sec:intro}
Spin glasses are disordered magnetic alloys~\cite{mydosh:93,young:98} that for many years now have fascinated both theoretical and experimental physicists~\cite{mezard:87,charbonneau:23,dahlberg:25}. Much of the interest in this subject originates in the many previously unheard-of features exhibited by the mean-field solution~\cite{parisi:79,parisi:80,parisi:83,mezard:84}, 
which have made these models relevant for biology, econophysics or computer science, among other fields~\cite{parisi:23}. We shall refer to the Parisi mean-field solution and to its consequences as the Replica-Symmetry Breaking (RSB) theory, which has grown to be an unavoidable paradigm for systems of high complexity. 

Establishing the mathematical correctness of RSB for the prototypical mean-field system ---the so-called Sherrington-Kirkpatrick model--- has been a long \emph{tour de force}~\cite{guerra:02,guerra:03,talagrand:06,panchenko:13}. Eventually it was proven that the Parisi solution is indeed correct for the Sherrington-Kirkpatrick model~\cite{talagrand:06}. When considering finite-dimensional, non-mean-field systems, most researchers believe that RSB provides an accurate description of spin glasses for space dimensions $D>D_{\text{u}}$ ---where $D_\text{u}=6$ is the  upper critical dimension in the absence of magnetic field--- although, missing a rigorous proof, some authors do not share the general view~\cite{fisher:88b,newman:25}. 

The discussion is centred around the spin-glass order parameter, the spin overlap $q$, and \df, the ---supposedly fractal--- dimension of the interfaces separating spatial regions where $q$ takes  different values: in a system of linear size $L$, the number of lattice links crossed by an interface scale as $L^{\df}$. The main open questions are about the probability $P(q=0)$ of finding states with $q=0$ and about whether or not $\df<D$. These questions have to be addressed in the $L\to\infty$ thermodynamic limit. If this limit is taken for a system in thermal equilibrium, the exact value of the temperature should not be relevant provided that $T<T_{\text{c}}$, where $T_\text{c}$ is the critical temperature that separates the paramagnetic phase  from the low-temperature spin-glass phase. 

There are (at least) four schools of thought regarding $P(q=0)$ and $\df$. The first possibility ---the RSB approach---  is that the excitations that lead to $q=0$ have a free-energy cost of $\mathcal{O}(1)$ and therefore $P(q=0)>0$ for all $T>0$. These excitations are space filling, meaning $\df=D$.

A second and somehow ``opposite'' possibility ---known as the droplet approach (DM)~\cite{mcmillan:84,bray:87,fisher:88b}--- is that only two pure states, connected by spin-inversion symmetry, exist and, similarly to what happens in a ferromagnet, all excitations are mixtures of these two. Excitations with $q=0$ have an excess energy that diverges with $L$, and their surface fractal dimension is $\df < D$.  This scenario leads to a trivial overlap distribution,  
$P(q)=q_\text{EA} \delta(q^2-q_\text{EA}^2)$, where $q_\text{EA}$ is the Edwards-Anderson order parameter. 

The two remaining scenarios were proposed later. The chaotic-pairs (CP) picture~\cite{newman:92,newman:96b,newman:97,newman:98} originated from the chaotic size dependence that jeopardizes naive attempts to take the thermodynamic limit. In fact, the metastate, a probability distribution over states, was introduced~\cite{aizenman:90,newman:92} in the hope that it will have a smoother $L\to\infty$ limit. The CP scenario predicts that excitations with $q=0$ have an energy cost that diverges as $L$ grows, so that  $P(q=0)$ is expected to vanish in the thermodynamic limit, and are space filling (hence $D=\df$). 

The trivial-non-trivial (TNT) picture~\cite{krzakala:00,palassini:00} starts with the observation that all numerical simulations at low 
temperatures indicate that $P(q\approx 0)$ is essentially $L$ independent and positive, as predicted by RSB; this $P(q=0)>0$ accounts for the 
``non-trivial'' in TNT. Then, these authors wonder about possible discrepancies with the RSB solution in $D=3$, motivated by the results 
of their numerical simulations that suggested that $\df<D$ ---this is the ``trivial'' part in TNT---, at odds with the RSB picture.

The  predictions from the four scenarios were nicely summarized in Ref.~\cite{newman:22b}:
\begin{itemize}
    \item RSB and CP: excitations leading to $q=0$ are space filling  (\emph{i.e.}, $\df=D$).
    \item RSB and TNT: excitations leading to $q=0$ have a free-energy cost of order 1, \emph{i.e.}, $P(q=0)>0$ for $T>0$.
\end{itemize}
It is worth stressing that, although it was not clear from the outset,  the metastate approach can be reconciled with RSB~\cite{read:14}. In fact, there are reasons to suspect that only DM, TNT, CP and RSB are mathematically consistent scenarios in the metastate framework~\cite{arguin:15,newman:24}. Furthermore, a numerical simulation using the Aizenman-Wher metastate seems to have erased DM from the list of scenarios that might be realized in $D=3$~\cite{billoire:17}. 

The above discussion raises the question: which of the above four scenarios is realized in finite-dimensional spin glasses? A sensible
way of answering our question must consider (typically three-dimensional) experimental data. Data from spin-glass materials such as CuMn is crucial, as it opens some interesting windows, but it does 
not reveal the full picture. The problem is that experimental spin glasses never reach thermal equilibrium at temperatures 
$T<T_\text{c}$, no matter how  long they are allowed to relax under stable conditions.  Although it is possible to use these 
relaxations to learn about equilibrium features~\cite{dahlberg:25}, these  lessons only apply to systems of limited size ($\sim 200$ lattice 
spacings, the largest spin-glass coherence length attainable nowadays \cite{zhai:19}, even in single-crystal samples).

If experiments do not suffice to close the debate, another possible avenue relies on analytical arguments~\cite{newman:25}. While this is an important approach, at present this analysis is not powerful enough 
to distinguish what happens to the equilibrium system in space dimensions below and above the upper critical dimension $D_{\text{u}}=6$.

The main way to answer our question that remains open is through numerical simulations, whose main limitation is in the modest size for which one is able to attain equilibrium. Here we will try and push these 
techniques. We shall mainly analyze the geometry of the excitations yielding $q\approx 0$. Our results have straightforward relevance 
to the analysis of experimental results: although close to $T_\text{c}$ coherence lengths as large as $\sim 200$ lattice spacings are reached \cite{zhai:19}, at 
low temperatures the experimentally achievable coherence length is directly comparable to the lattice size of our 
simulations~\cite{orbach-janus:24}. Moreover, when the lattice size has this magnitude, we always find values of $P(q=0)>0$. In fact, we shall consider $P(q)$ only to assess that we have reached temperatures low enough to make contact with the previous body of work that we discuss next.

Up to now, low-energy excitations with $q\approx 0$ have been commonly studied by comparing the ground state of instances of the system before and after a perturbation to the couplings of the Hamiltonian. Working at $T=0$ has some obvious advantages: first, sampling is exact, because only 
the ground state survives the $T\to 0$ limit; second, the energy ---which is a computable quantity--- and the free energy ---a more interesting 
but much more elusive quantity--- coincide at $T=0$. 

The issue with the $T=0$ approach is that one has to decide what a ``natural'' perturbation is. The goal is inducing in the new ground state features similar to those characterizing the true $T>0$ excitations. 

Proposals include perturbation to a single bond~\cite{shen:24}, or to the whole system~\cite{marinari:01, palassini:00}. Both methods force an $\mathcal{O}(1)$ variation in energy: if some of the perturbed ground states have a small overlap with respect to the initial one, one may 
hope that those configurations will contribute to $P(q=0)$ at low temperatures. In addition, when the perturbation is applied by changing 
the  boundary conditions~\cite{marinari:00d}, the energy variation is much smaller than $\mathcal{O}(L^D)$ but is not guaranteed to be of $\mathcal{O}(1)$. 

Let us stress the two main physical hypotheses underlying the strategy of simulating directly at $T=0$. The first crucial assumption is that the limit of zero temperature is smooth. To be precise, one assumes that the two limits $T\to 0 $ and $L\to\infty$ commute. We shall comment more on this assumption below. The second assumption is formulated in terms of  the link overlap $\ql$ defined in Sec.~\ref{sec:model_obs}, which measures the 
density of lattice bonds that are \emph{not} crossed by the interface of an excitation. Then, in  $T=0$ simulations,  one is implicitly  relying on overlap equivalence~\cite{contucci:06}, which assumes that $\ql$ becomes a one-to-one function of the squared spin-overlap $q^2$ when $L\to\infty$. The overlap-equivalence property is trivial in the Sherrington-Kirkpatrick model ---for which $\ql$ is \emph{defined} as $q^2$---, but in finite dimensions it remains  controversial. In fact, TNT supporters claim that $\ql$ becomes a constant function, while careful numerical simulations at finite temperature indicate that the dependence on $q^2$ is indeed one to one~\cite{janus:10}. 

Let us now  discuss the main conclusions reached by the $T=0$ simulations. Shen \emph{et al.}~\cite{shen:24} use very large lattices in 
$D=2$, but in $D=3$ they reach lattices up to $L=12$, while more than twenty years ago Palassini and Young reached $L=8$~\cite{palassini:00} 
and Marinari and Parisi went up to $L=14$~\cite{marinari:01}. When comparing these works, the first question that arises is concerned with the equivalence, 
or lack thereof, of the single-bond~\cite{shen:24} and the whole-system~\cite{palassini:00,marinari:01} perturbation methods. We answer the question in the negative, as we explain next.

The probability density function for the surface $\mathcal{A}$ of the excitations generated through single-bond perturbation turns out to be 
a barely normalizable power law  $\rho(\mathcal{A})\sim 1/\mathcal{A}^{1+\kappa_\textsc{S}}$ with a cutoff at 
$\mathcal{A}\sim L^{d_\textsc{S}}$. The finite-size scaling analysis of Ref.~\cite{shen:24} finds for the exponents $\kappa_\textsc{S}=0.159(5)$ and $d_\textsc{S}=2.76(2)[15]$, where 
the first error is statistical and the second  error, between square brackets, accounts for systematic  effects. These findings imply that the median $\mathcal{A}$  from single-bond  perturbation is of order 1. In fact, percentile 99 of $\mathcal{A}$ also scales as $L^0$ and excitations with surfaces of order $L^{d_\textsc{S}}$ are rare events that arise with probability 
$\sim 1/L^{d_\textsc{S}\kappa_\textsc{S}}$. On the other hand, rare-event analysis does not play any role when dealing  with bulk perturbations~\cite{palassini:00,marinari:01}. 
If one insists in comparing the two  perturbation methods 
through the first moment of their respective probability distributions, the cutoff $\mathcal{A}\sim L^{d_\textsc{S}}$ implies 
that the average surface from  single-bond  perturbation scales as $L^{d_\textsc{S}(1-\kappa_\textsc{S})}$ with $d_\textsc{S}(1-\kappa_\textsc{S})\approx 2.32(14)$, which 
is 1.9 standard deviations below the corresponding result from bulk perturbations in small systems $\df=2.58(2)$~\cite{palassini:00} (one gets $\df=3$ 
from bulk perturbations when larger systems are considered~\cite{marinari:01}). From this argument, as well as from the expanded discussion in Sec.~\ref{sec:final}, which relates this topic  to overlap equivalence, we conclude that only the bulk-perturbation method relates  naturally with the excitations that 
appear in a  spin glass at $T>0$.

The reader will have noticed, however, that contrasting conclusions have been reached with the bulk-perturbation method: while 
Ref.~\cite{palassini:00} favors non-space-filling surfaces, Ref.~\cite{marinari:01} claims that $\df=3$ in $D=3$ systems. Compared to Ref.~\cite{palassini:00}, a crucial novelty in the analysis of Ref.~\cite{marinari:01} is that they ensured that the surface and the volume of the \emph{same} excitation were being confronted by conditioning the probability distribution to the value of the spin overlap~$q$. Based on this, as well as
on the extended range of system sizes, it was concluded that $\df=D$, although significant corrections to the leading scaling are present in 
the data~\cite{marinari:01}. When these considerations were taken into account by the authors of Ref.~\cite{palassini:00} in a later paper~\cite{palassini:03}, they found results compatible with the scenario of space-filling surfaces (although they preferred the non-space-filling interpretation of their data).

Here, we adopt a more straightforward approach to the problem by directly studying the geometry of the interfaces for excitations with 
$q\approx 0$, on data collected from low-temperature configurations corresponding to systems of linear sizes up to $L=18$ (low temperature means here  $T=0.2\approx  0.21  T_\text{c}$).  In other words, we allow the system to select its own excitations. The same strategy was followed, 
down to our same temperature, in  Ref.~\cite{katzgraber:01}, which was however limited to smaller systems, with $L\leq 8$. We use the so-called Parisi-Toulouse (PaT) scaling to assess that we 
have reached the low-temperature regime of interest ---see Ref.~\cite{parisi:80e} and our discussion in Sec.~\ref{sec:num_results}; let us stress that  PaT 
scaling provides strong support for the hypothesis of smoothness in the $T\to 0$ limit.  In this way, when we restrict our analysis to the sizes explored 
by Ref.~\cite{palassini:00,katzgraber:01} we find results consistent with their $\df$. However, as we allow $L$ to grow, the estimate of $\df$ increases as well. 
In fact, the simplest interpretation of our results is that $\df=3$, in agreement with Ref.~\cite{marinari:01}. Our final results are also
numerically consistent with the outcome from  a non-equilibrium simulation~\cite{marinari:02}.

The reader will note that we are simulating systems larger than previously thought possible for such a low temperature as $T=0.2$ (at 
least without relying on unconventional hardware~\cite{janus:14}). Indeed, we are fully equilibrating down to $T=0.2$ lattices with size $L=16$ and we manage to obtain useful results for $L=18$ systems. The crucial ingredient for this success is Houdayer's cluster move~\cite{houdayer:01}, which
was originally introduced for the simulation of two-dimensional spin glasses. Ref.~\cite{zhu:15} suggested that the move could also be useful for $D=3$ systems, but only presented equilibrated systems up to size $L= 12$, and that for temperatures $T\geq 0.42$~\cite{zhu:15}. We have significantly improved over~\cite{zhu:15}, both in system size and in minimal temperature, by introducing modifications to the simulation algorithm 
in a procedure we call Parallel Tempering enhanced with Houdayer moves and with the entropic Reservoir, shortened to PTHR, which we describe in a
separate work \cite{chilin:26b}.

The rest of this work is organized as follows. Sec.~\ref{sec:model_obs} describes the Hamiltonian and the observables that we have used to investigate the geometric 
properties of the interfaces, whose physical meaning is briefly discussed. 
At the end of this section we give a brief description of our numerical simulations, and a more detailed report is in  Appendix~\ref{sec:appendice_tecnico}. In Sec.~\ref{sec:num_results} we show the data analysis that led to our results, including the examination of PaT scaling, the 
estimate of $\df$ for $q\approx 0$ excitations, the extrapolation to large system sizes, and the extrapolation for $T \to 0$. In Sec.~\ref{subsubsec:E-limit} we also extrapolate to $T=0$ our estimates for the energy density that have previously been extrapolated to the thermodynamic limit. This gives us the occasion to compare ---favourably--- with previous work that took the limits in the reverse order.
We present our conclusions in Sec.~\ref{sec:final}, where we dedicate a separate subsection, Sec.~\ref{subsec:otherworks}, to a detailed comparison  with previous work.

\section{Model and observables}\label{sec:model_obs}

In this section we give the basic definitions of our two main observables and discuss their physical significance. A brief description of our numerical simulations, expanded in Appendix~\ref{sec:appendice_tecnico}, is provided at the end of the section.

We study the three-dimensional Edwards-Anderson spin glass (EA3D)~\cite{edwards:75, edwards:76}:
\begin{equation} \label{eq:hamiltonian}    
\mathcal{H} \left\{ \boldsymbol{\sigma} \right\} = -\sum_{ \langle i, j \rangle } J_{ij}\, \sigma_i\, \sigma_j\,.
\end{equation}
The couplings $J_{ij}$ connect all nearest-neighbour pairs, as denoted by the $\langle i,j\rangle$, in a $D=3$ simple-cubic lattice of side $L$ with periodic boundary conditions. The $V=L^3$ Ising-like spins $\sigma_i=\pm 1$ are our dynamical variables. The $J_{ij}$ are often chosen as binary variables as well, but for low temperatures this version of the model can lead to unwanted artifacts~\cite{janus:10}. Instead, we extract the $J_{ij}$ from a Gaussian distribution with zero mean and unit variance. With this choice of couplings, $T_\text{c}=0.95(4)$~\cite{marinari:98d}; see Ref. ~\cite{janus:13} for a determination of the critical exponents and other universal quantities. 
The $J_{ij}$ are quenched, \emph{i.e.}, fixed at the beginning of the simulation, each instance of the $\{J_{ij}\}$ defining a sample. After obtaining the single-sample thermal average $\tavg{\mathcal{O}}$ for an observable $\mathcal{O}$, a second average over the disorder ---\emph{i.e.}, over the $\{J_{ij}\}$--- has to be computed and is denoted
by $\davg{ \mathcal{O}}$.

Many observable quantities are best defined by introducing real replicas. These are copies of the system that share the same  $\{J_{ij}\}$ but evolve independently. The most important example is the (spin) overlap 
\begin{align} \label{eq:tot_ovlp}
   q&=\frac{1}{V} \sum_{i=1}^{V} q_i\,, & q_i &= \sigma_i^{(a)} \sigma_i^{(b)}\,,
\end{align}
where the superindices $(a)$ and $(b)$ denote different real replicas.

The spin overlap, which is the order parameter for the spin-glass transition, is straightforwardly related to the Hamming distance $d_\textsc{H}$, the number of sites in which $\sigma_i^{(a)} = -\sigma_i^{(b)}$, as $q=1-2 d_\text{H}/V$. Because any of the two replicas can be globally flipped without changing the energy, it is equally probable to find either $q$ or $-q$. 
In fact, we define the probability 
\begin{equation}
    P(q=c) = \davg{ \delta_{q, c} } \,,
\end{equation}
where $\delta_{q,c}$ is the Kronecker delta symbol, which is indeed an even function. Because of this it can be appropriate to focus on $q^2$ rather than on $q$. In fact, the largest meaningful Hamming distance for the configurations of two replicas is $d_\text{H}=V/2$, which corresponds to $q=0$.

A $q=0$ excitation can also be built in a trivial way. It suffices to select a domain, for example a sphere, containing half of the sites of the lattice and reversing all the internal spins to have $q=0$ for the overlap between the initial and final configurations. The surface-to-volume ratio of this trivial excitation goes to $0$ as $1/L$. In order to distinguish trivial from non-trivial excitations we introduce the \textit{link overlap} 
\begin{equation}\label{eq:link_ovlp}
    q_\text{l} = \frac{1}{3  V} \sum_{\langle i, j \rangle} \sigma_i^{(a)} \sigma_i^{(b)} \sigma_j^{(a)} \sigma_j^{(b)} \,.
\end{equation}
A lattice bond $ij$ where $\sigma^{(a)}_i \sigma^{(a)}_j=\sigma^{(b)}_i \sigma^{(b)}_j$ gives a positive contribution to $q_\text{l}$. Hence, only the bonds that are traversed by the surface(s) of the excitation give a negative contribution to $\ql$. This is why $1-\ql$ effectively acts as a measure of the size of the interface between regions with $q_i=1$ and $q_i=-1$. We can then write the surface-to-volume ratio of the excitation in terms of spin~\eqref{eq:tot_ovlp} and link overlap~\eqref{eq:link_ovlp}:
\begin{equation}
    \frac{(1-\ql)DV/2}{(1-q)V/2} \sim \frac{L^{\df}}{L^D}=\frac{1}{L^{D-\df}} \, ,
\end{equation}
where, as we discuss in the introduction, $\df$ is the exponent that regulates the scaling of the interface size.

We focus now on the $q=0$ excitations, by considering the link overlap conditioned to $q=0$. To do so, we need to define the conditional  expectation value for $\ql$ (for any other observable the definition would be analogous).  A natural choice is
\begin{equation}
\E(\ql|q^2=c)=\overline{\langle\, \ql \,\delta_{q^2,c}\, \rangle}/{\overline{\langle \delta_{q^2,c} \rangle}}\,,
\end{equation}
which studies $q^2$ at its microscopic level of discretization, $q^2=(i/V)^2$ with $i=0,2,4,\ldots V\,$. It has long been recognized that some form of coarse graining significantly improves the statistical accuracy of the result~\cite{janus:10}. Our solution is simply to coarse grain to the microscopic levels of an $L=4$ system:
\begin{equation}\label{eq:conditional-expectation}
    \E(\ql|q^2=\Tilde{c}_i)=\frac{\overline{\langle\, \ql \, \chi^{(i)}(q^2)}\, \rangle}{\overline{\langle \chi^{(i)}(q^2)\rangle}}\,,\quad
    \Tilde{c}_i = \frac{\overline{\langle\, q^2 \, \chi^{(i)}(q^2)}\, \rangle}{\overline{\langle \chi^{(i)}(q^2)\rangle}}\,,
\end{equation}
where $\chi^{(i)}(q^2)$ is an indicator function for the coarse graining, 
\begin{equation}
    \chi^{(i)}(q^2)= 
\begin{cases}
    1\hspace{0.5cm} \text{if } \sqrt{q^2} \in [i/4^3, (i+2)/4^3) \\
    0\hspace{0.5cm} \text{otherwise } 
\end{cases}\,\quad i=0,2,4,\ldots,4^3-2,
\end{equation}
with the exception of $i=4^3-2$, for which we take a compact interval $\sqrt{q^2}\in[1-2/4^3,1]$. A more rigorous notation would have been $\E(q_{\text{l}}|\chi^{(i)}(q)=1)$, but we prefer $\E(\ql|q^2=\Tilde{c}_i)$, which is in some sense more direct. Similarly, the conditional variance of $\ql$ is
\begin{equation}\label{eq:def-conditional-var}
    \text{Var}(\ql | q^2=\tilde{c}_i)=\E(\ql^2|q^2=\tilde{c}_i)-[\E(\ql | q^2=\tilde{c}_i)]^2\,.
\end{equation}
The standard variance and the standard expectation value  are straightforwardly related to the conditional ones through the following sum rules:\footnote{The sum rules are specific to the  chosen coarse graining in $q^2$, but always have the same overall structure~\cite{janus:10}.}
\begin{eqnarray}\label{eq:sum-rule-1}
    \overline{\langle \ql\rangle}&=&\sum_{i=0}^{4^3-2}\,{\overline{\langle \chi^{(i)}(q^2)\rangle}}\, \E(\ql|q^2=\Tilde{c}_i)\,,\\
    \text{Var}(\ql)&=&\sum_{i=0}^{4^3-2}\,{\overline{\langle \chi^{(i)}(q^2)\rangle}}\, \Big[\text{Var}(\ql|q^2=\Tilde{c}_i)\ +\  \big(\,\E(\ql|q^2=\Tilde{c}_i)-\overline{\langle \ql\rangle}\,\big)^2\,\Big]\,.\label{eq:sum-rule-2}
\end{eqnarray}
Both the TNT and the DM pictures imply that the conditional expectation of $\ql$ will vanish at large $L$ and $T=0$ as
\begin{equation}\label{eq:the_crux_of_the_matter}
1-\E(\ql | q^2=0) \propto \frac{1}{L^{D-\df}}\,.
\end{equation}
Instead, both the CP and the RSB pictures predict that $D=d_{\text{f}}$, which does not necessarily imply that $\E(\ql | q^2=0)<1$ in the  thermodynamic limit because one could still have a marginal behaviour such as, for instance, $[1-\E(\ql | q^2=0)]\propto 1/\log L$. For models where RSB provides the exact solution, however, one has $\E(\ql | q^2=0)<\davg{\ql}$ in the thermodynamic limit for any $T<T_{\text{c}}$: this is  trivial to show in the Sherrington-Kirkpatrick model ---for which $\ql=q^2$--- and the same conclusion was reached in numerical  simulations of mean-field models with finite connectivity~\cite{fernandez:09f}. Furthermore,  having excitations with space-filling  surfaces  is crucial for the overlap-equivalence property~\cite{contucci:06}, which predicts that $\E(\ql|q^2=c)$ will be a one-to-one function of $c$.

We conclude this section by considering the Schwinger-Dyson-Young (SDY) observable for Ising spins on a simple-cubic lattice with Gaussian-distributed couplings:
\begin{equation}\label{eq:SDY-def}
   {\cal S} =  \frac{\cal H}{3 V} + \frac{(1-q_\text{l})}{T}\,.
\end{equation}
Then, but \emph{only in thermal equilibrium}, we have a Schwinger-Dyson-like identity~\cite{bray:80c}\begin{equation}\label{eq:SDY}
   \overline{\langle {\cal S}\rangle} =0 \,. 
\end{equation}
Eq.~\eqref{eq:SDY} can, therefore, be used as a strict 
thermalization test~\cite{katzgraber:01}.

 We ran simulations for linear sizes $L=4, 6, 8, 10, 12, 14, 16$, and $18$. Data for $L\leq 12$ was produced with standard Parallel Tempering simulations~\cite{hukushima:96,marinari:98b}. For $L=14, 16$, and  18, we employed  the enhanced PTHR method, explained in Ref.~\cite{chilin:26b}. The lowest temperature we simulated for $L\geq 8$ was $T_\text{min}=0.2$, while it was possible to reach $T_\text{min}=0.05$ for $L=4, 6$. We then computed the $T \to 0$ limit from our data.

In order to convince ourselves that thermal equilibrium had been reached, we checked that $\overline{\langle \mathcal{S}\rangle}$ was compatible with zero for the data at all system sizes, see Eq.~\eqref{eq:SDY}. This criterion was fully satisfied for $L\leq16$, while $L=18$ was only nearly equilibrated. Data for this larger size, however, is still useful thanks to the overlap-equivalence property, as explained in detail in Sec.~\ref{subsec:useof18}.

Further details on the simulations are provided in Appendix~\ref{sec:appendice_tecnico}. 

\section{Numerical results}\label{sec:num_results}

In this section we shall address two questions. Our main goal is obtaining an estimate of the fractal dimension $\df$ through the scaling in Eq.~\eqref{eq:the_crux_of_the_matter}. A second  question arose, however, when comparing with the $T=0$ results in  Ref.~\cite{marinari:01}: when lattice sizes are comparable, their  $T=0$ results turned out to be almost compatible with our $T=0.2$ data; see the discussion in Sec.~\ref{subsec:TfiniteT0}. This unexpected coincidence begged for an explanation, and we  found one. Our reasoning stands on two legs: (i) overlap equivalence, and (ii) the fact that we  are indeed exploring the physics of asymptotically low temperatures.  This is why we have decided to start the presentation of our numerical results by showing in  Sec.~\ref{subsec:PaT} that our $P(q)$, verifies PaT scaling and that, as a consequence, it really reflects the low-temperature physics.

Based on this firmer ground, we start our study by taking the $q^2\to 0$ limit for $\qlqsq$  in Sec.~\ref{subsec:qsq0}. Next, we take the $L\to\infty$ and $T\to 0$ limits in Sec.~\ref{subsec:limits}. Further discussions will be found in Sec.~\ref{subsec:otherworks}, where we compare our results  with previous work and present our proposed explanation for the surprising similarity between our $T>0$ results and those for $T=0$ in Ref.~\cite{marinari:01}. 

Nevertheless, the reader may find it useful that we anticipate here our main result. Our values of $\qlqz$ are extrapolated to the thermodynamic limit first. Then, this limit is extrapolated to zero temperature. This final extrapolation is 
\begin{equation}
  \E \left (\davg{q_\text{l}} | q^2=0 \right) \bigg\rvert_{ T\to 0, L\to\infty}= 0.8085 (18)[212]\,,
\end{equation}
where the first error bar is statistical, and the second error (in square brackets) accounts for systematic effects. Since this value is clearly smaller than 1, the most economical interpretation of our results is $\df=3$.

\subsection[Parisi-Toulouse scaling for $P(q)$ at low temperatures]{Parisi-Toulouse scaling for \boldmath $P(q)$ at low temperatures}\label{subsec:PaT}

\begin{table}[b]
\centering
\begin{tabular}{ r l l l l l }
\hline \hline
\multicolumn{1}{c}{$L$} & \multicolumn{1}{c}{$T_\text{min}$} & \multicolumn{1}{c}{$\text{Prob}[q^2<0.01]$}&
\multicolumn{1}{c}{$\mathcal{Q}_1$}&
\multicolumn{1}{c}{$\mathcal{Q}_2$}&
\multicolumn{1}{c}{$\mathcal{Q}_3$}
\\ \hline
4  & $0.05$ & $0.00231 (15)$ & $0.8113 (16)$ & $0.9261 (5)$  & $0.9667 (1)$ \\
6  & $0.05$ & $0.00224 (17)$ & $0.8049 (9) $ & $0.9206 (2) $ & $0.9607 (3)$ \\
8  & $0.2$  & $0.0092 (5)$   & $0.812 (2)$ & $0.9067 (4) $  & $0.9459 (5)$ \\
10 & $0.2$  & $0.0097 (12)$  & $0.806 (4)  $ & $0.9007 (12)  $ & $0.9390 (6)$  \\
12 & $0.2$  &$ 0.0108 (14)$  & $0.798  (5) $ & $0.8925 (14) $  & $0.9295 (5)$  \\
14 & $0.2$  &$ 0.0077 (10)$  & $0.798 (4) $  & $0.8853 (10)$ & $0.9211 (5)$  \\
16 & $0.2$  & $0.0093 (10) $ & $0.796 (4) $  & $0.8811 (10) $ & $0.9153 (5)$  \\ \hline \hline
\end{tabular}
\caption{\label{tab:pq_per_size}
      $\text{Prob}[q^2<0.01]$, Eq.~\eqref{eq:prob-q2-def}, as well as the first and third quartiles $\mathcal{Q}_{1,3}$ and the median $\mathcal{Q}_2$ of the cumulative $\mathcal{F}(q^2_\text{max})$ at the minimum temperatures of our simulations, for the sizes that we managed to equilibrate fully $[\mathcal{F}(q^2_\text{max}=\mathcal Q_i)=i/4]$. Errors are computed using a bootstrap method.}
\end{table}

\begin{figure}
    \centering
    \includegraphics[width=\linewidth]{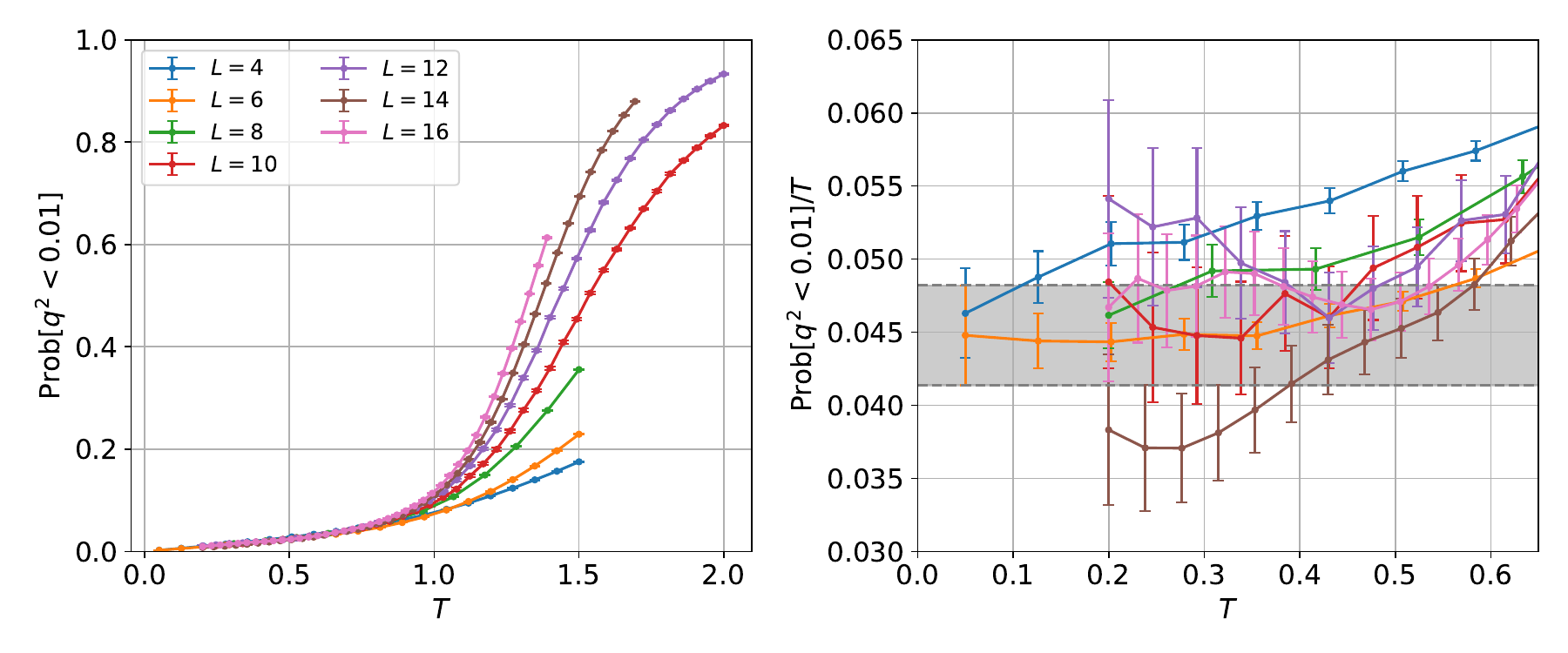}
    \caption{\label{fig:PaTscaling}
        Parisi-Toulouse scaling. Temperature and size dependence for the cumulative probability $\text{Prob}[q^2<0.01]$, see Eq.~\eqref{eq:prob-q2-def}, which provides a numerically stable estimate of $P(q=0)$. \textit{Left:} $\text{Prob}[q^2<0.01]$ versus  temperature, as computed for all our fully equilibrated lattice sizes. \textit{Right:} Closeup of the $T \in [0, 0.65]$ region, where we show $\text{Prob}[q^2<0.01]/T$; PaT scaling predicts that $\text{Prob}[q^2<0.01]/T$ should be temperature independent. The grey horizontal band corresponds to $\text{Prob}[q^2<0.01;L=6, T=0.05]$ plus or minus its error. Excepting $L=4$, all our data is compatible with this band if $T<0.5$. The key is common to the two panels.
    }
\end{figure}

The fact that, at low temperatures, $P(q=0)$ does not show any tendency to vanish as $L$ grows has been known for a long time. Let us recall that numerical results for $P(q)$ close to $q=0$ have been reported both for Gaussian couplings ---on lattice sizes up to $L=16$, using the APE100 special-purpose computer~\cite{marinari:00}--- and for  binary couplings ---on sizes up to $L=32$ that were equilibrated using the Janus computer~\cite{janus:10}. However, lattice size is mainly limited by the minimal temperature in the simulation. For instance, $L=32$ could be thermalized for temperatures  $T\geq T_\text{min}\approx 0.64 T_\text{c}$ in~\cite{janus:10}. Ref.~\cite{katzgraber:01} reached much lower temperatures, $T_\text{min}=0.2\approx 0.21 T_\text{c}$ with Gaussian-distributed couplings, but was limited to $L\leq 8$. Here, we also have $T_\text{min}=0.2$ but
we are able to equilibrate lattices up to $L=16$.

Since the  matter of primary interest here is $P(q=0)$, one could be tempted to consider this quantity directly, which would be a bad idea because configurations with exactly $q=0$ are a pretty uncommon event. In other words, the data for $q=0$ is noisy. To cure this problem, we prefer to consider cumulative probability functions
\begin{equation}\label{eq:prob-q2-def}
    \mathcal{F}(q_{\text{max}}^2)\equiv \text{Prob}[q^2<q_\text{max}^2] = \int_{-q_\text{max}}^{q_\text{max}} \dd q \ P(q) \,.
\end{equation}
Notice that $\mathcal{F}(q_\text{max}^2)/(2q_\text{max})$  should approach $P(q=0)$ as $q_\text{max}$ goes to zero. We found no difference ---within error--- for $q_\text{max}^2 = 0.01, 0.1$; hence, we decided to use $q_\text{max}=0.1$ throughout this section.

As Fig.~\ref{fig:PaTscaling} displays and Table \ref{tab:pq_per_size} shows in greater detail for our minimal temperature, at any fixed 
temperature $T<0.6$, $\text{Prob}[q^2<0.01]$ does not show any measurable size dependence. Furthermore, quartiles $\mathcal Q_1$, $\mathcal Q_2$ and $\mathcal Q_3$ of the 
cumulative distribution~\eqref{eq:prob-q2-def} are slightly decreasing functions of $L$, which indicates that probability is (if anything) 
concentrating on the small-$q$ region as $L$ grows.

Indeed, the most notable dependence in Fig.~\ref{fig:PaTscaling} is on \emph{temperature}, which immediately calls the question: how low does $T$ need to be
to reach the  low-$T$ regime that interests us here? To answer this question, we recall that ---if the temperature is low 
enough---  $P(q\approx 0)$ should be linear in $T$. This temperature dependence directly follows from  PaT scaling~\cite{parisi:80e}.  We reproduce the argument in a separate paragraph below, because Ref.~\cite{parisi:80e} does not directly address $P(q)$. As for the numerical data,  PaT scaling was checked with Gaussian couplings and down to $T_\text{min}=0.2$ on lattices $L\leq 8$, see Fig. 5 of Ref.~\cite{katzgraber:01}. With our
larger system sizes, we confirm
in the right panel of Fig.~\ref{fig:PaTscaling} that temperatures lower than $T<0.5$ meet the 
PaT condition. 

\subsubsection[From the PaT assumption to the temperature dependence of \texorpdfstring{$P(q)$}{P(q)}]{From the PaT assumption \boldmath to the temperature dependence of \texorpdfstring{$P(q)$}{P(q)}}
PaT scaling~\cite{parisi:80e} is a low-temperature behaviour predicted by RSB. The main assumption is that, for $T\sim 0$, the entropy $S_\alpha$ of any state $\alpha$ does not vary much with temperature. 
We can then approximately write the thermodynamic weights of the states that contribute to the partition function as 
\begin{equation}
    w_\alpha (T) \sim c_\alpha \ee^{- [E_\alpha(T) -E_\text{min}(T)]/T} \,,
\end{equation}
where the normalizing factors $c_\alpha$ are (almost) temperature independent and $E_\text{min}(T)$ is the energy of the state that minimizes the free energy at temperature $T$. Let us collect all the contributions from states having energies $E_\alpha$ and 
overlap $q=q_{\alpha}$ with the state of minimal free energy at temperature $T$ in $\mathcal{N}(E,q;T)=\sum_{\alpha} c_\alpha \delta(q-q_{\alpha})\,\delta(E - E_\alpha(T))$. RSB has a very unusual prediction for $\mathcal{N}(E_\text{min}+\epsilon,q;T)$, where $\epsilon$ is a positive number of order one: if one first takes  the thermodynamic limit and only afterwards the
limit of $T,\epsilon\to 0$, a positive quantity $\mathcal{N}(E_\text{GS},q;T=0)$ remains.
Then, we have for $P(q; T)$ at low temperatures:
\begin{align}\nonumber
    P(q; T) & = \int_{E_\text{min}}^\infty dE \ \mathcal{N}(E, q;T) \ e^{- (E-E_\text{min})/T} \, \\\nonumber
            & \simeq  \int_{E_\text{min}}^\infty dE \ \mathcal{N}(E_\text{min}, q;T) \ e^{-(E-E_\text{min})/T} \\
            & = \mathcal{N}(E_\text{min}, q;T) \,T\ \simeq\  \mathcal{N}(E_\text{GS}, q;T=0)\,  T\,.
\end{align}

\subsection[Extrapolating $\E(q_\text{l}|q^2)$ to $q^2=0$]{Extrapolating \boldmath $\E(q_\text{l}|q^2)$ to $q^2=0$}\label{subsec:qsq0}

\begin{figure} 
    \includegraphics[width=\linewidth]{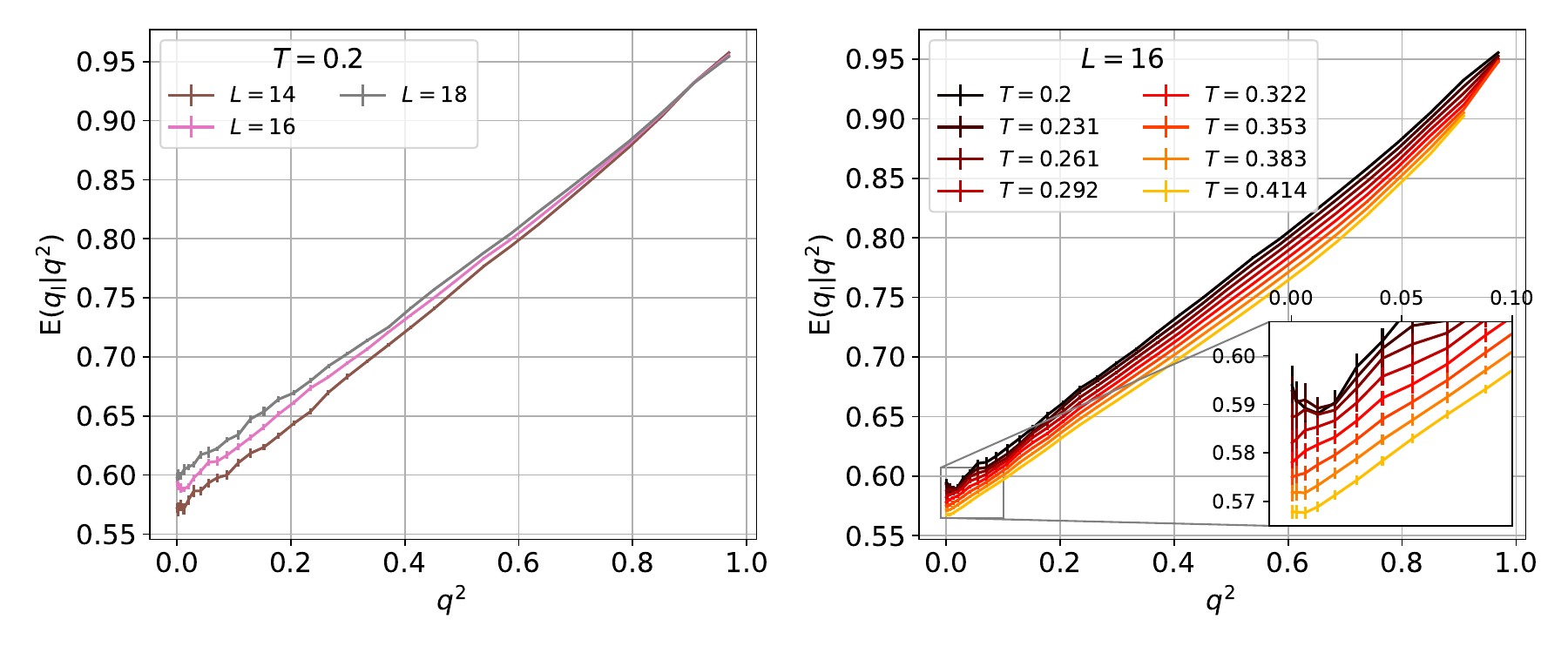}\par 
\caption{\label{fig:pq_qlqsq}
    {Link overlap as a function of} $q^2$. Plot of $\E(q_\text{l} | q^2)$, see Eq.~\eqref{eq:conditional-expectation}, as computed for the lowest temperatures and for the biggest sizes we have simulated. In both panels errors are computed using a jackknife method.
    \textit{Left:} estimate of $\E(q_\text{l} | q^2)$ for $L=14, 16, 18$ at $T=0.2$. 
    \textit{Right:} estimate of $\E(q_\text{l} | q^2)$ for $L=16$ at various temperatures. The inset shows a closeup of the highlighted box. 
}
\end{figure}

Our main goal here is to obtain the best possible extrapolation of $\E(q_\text{l} | q^2)$ to $q^2=0$, see  Eq.~\eqref{eq:link_ovlp}. We  need  $\E(q_\text{l} | q^2=0)$ as a function of $T$ for all our system sizes. Our raw data for $\E(q_\text{l} | q^2)$ is shown in
Fig.~\ref{fig:pq_qlqsq}. Probably, the reader will be surprised by the smallness of the statistical errors near $q^2=0$. Indeed, finding
a value of $q$ close to $0$ is very infrequent, particularly for low temperatures, as shown in 
Table~\ref{tab:pq_per_size} and in Fig.~\ref{fig:PaTscaling}, and discussed in Sec.~\ref{subsec:PaT}. 
Nevertheless, the statistical errors for $\E (\ql | q^2 )$ for $q\approx 0$ are in all cases below 1\%; see Fig.~\ref{fig:pq_qlqsq}, especially the inset in the right panel. 

The way out of the paradox lies in the numerical observation that the conditional variance for the link overlap, see Eq.~\eqref{eq:def-conditional-var},  vanishes when $L$ grows as
$\text{Var}(\ql | q^2)\propto 1/L^{D/2}$~\cite{contucci:06,janus:10}. This behaviour of the conditional variance implies
that $\ql$ becomes a deterministic (\emph{i.e.}, non-stochastic) function of $q^2$ for large systems and it is a prerequisite for overlap 
equivalence~\cite{contucci:06}. Notice as well that having $\mathrm{Var}(\ql | q^2)\to 0$ is not in contradiction with the TNT picture: TNT 
simply states that, in the $L\to\infty$ limit, the non-stochastic $\qlqsq$  is a trivial constant as $q^2$ varies.

\begin{figure}
    \centering
    \includegraphics[width=\linewidth]{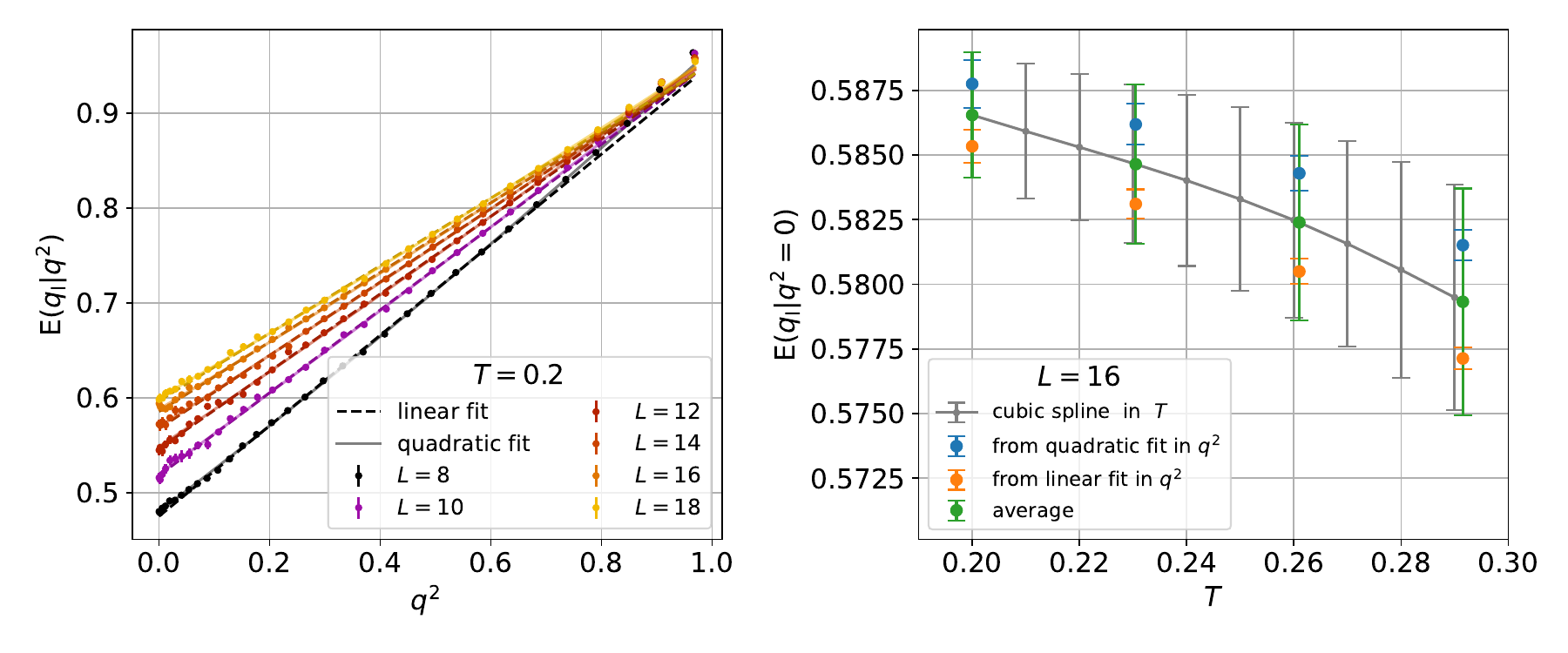}
    \caption{\label{fig:fits_and_interpols} 
        {Extrapolating} $\E(q_\text{l} | q^2)$ {to} $q^2=0$. Through the procedure explained in Sec.~\ref{subsec:qsq0} we get interpolated data points for equally spaced temperatures $0.2 \leq T < 0.3$.
        \textit{Left:} $\E(q_\text{l} | q^2)$ at $T=0.2$ for different linear sizes $L$. Dots represent the measurements, dashed lines are fits to a linear function, solid lines are fits to a quadratic function. In order to get cleaner results, fits have been restricted to 
        the intervals shown in Appendix~\ref{subsec:appendice_tecnico_q2_to_zero}.
        \textit{Right:} $\E(q_\text{l} | q^2 = 0)$ for $L=16$ at different temperatures $T$. Blue and orange dots are the values of the intercepts obtained from the fits. Green dots are the average between the two, as explained in the main text. The grey curve is the cubic-spline interpolation between the green dots.
    }
\end{figure}

However, although the small conditional variance protects us, visual inspection of the right panel of Fig.~\ref{fig:pq_qlqsq} tells us that 
statistical errors increase very significantly as $q^2\to 0$, where the behaviour of the numerical curve becomes somewhat irregular. This is why we have preferred to smooth our data by performing fits
of our numerical estimates of $\qlqsq$ to both linear $f(q^2)=a_0+a_1 q^2$ and quadratic $g(q^2)=b_0+b_1 q^2 + b_2 q^4$ functions, as exemplified in the left 
panel of Fig.~\ref{fig:fits_and_interpols}, see also Appendix~\ref{subsec:appendice_tecnico_q2_to_zero}. The resulting $a_0$ and $b_0$ intercepts for the four lowest temperatures of the $L=16$ data are shown 
in the right panel as the orange and blue points, respectively. The statistical errors for $a_0$ and $b_0$ that are displayed in Fig.~\ref{fig:fits_and_interpols}-right were computed using the jackknife method as explained in~\cite{yllanes:11}. However,
 the statistical errors for $a_0$ and $b_0$ turned out to be in all cases smaller than the difference $|a_0-b_0|$. We have therefore chosen  to use the mean between the two intercepts as $\E ( q_\text{l}| q^2=0 )=(a_0+b_0)/2$ (green dots in Fig.~\ref{fig:fits_and_interpols}, left panel) and the difference between the two $|a_0-b_0|$ as our error.

There is a final problem to be addressed, namely that the simulations for different sizes were not 
at the same temperatures ---see Table~\ref{tab:PTparameters}--- which complicates the comparison of data for different $L$. We solved this
problem using a cubic-spline interpolation of the data in the $T\in[0.2, 0.3)$ interval, 
eventually obtaining $\qlqsq$ as a continuous function of $T$.\footnote{We employed the so-called natural cubic spline~\cite{press:92}.\label{footnote:cubic-spline}}
As for the error bars, we interpolated them in temperature using the same procedure. The grey dots in the right panel of Fig.~\ref{fig:fits_and_interpols} were obtained following this procedure for $L=16$.

\subsection{The thermodynamic limit and the zero-temperature limit}\label{subsec:limits}

Once we have in our hands the extrapolations $\qlqz$ at all temperatures $0.2 \leq T <0.3$ for all sizes, our next goal is to reach the thermodynamic 
limit at fixed $T$. We do that by performing a fit in $L$ of $1-\qlqz$ and inferring the $L\to\infty$ value from the obtained function. As explained in Sec.~\ref{subsubsec:E-limit}, the  same strategy can be successfully applied to the energy density $\davg{\mathcal{H}/V}$.

For any finite $T$ one expects
\begin{equation}
0\ < \ \lim_{L\to\infty}\,\qlqz\ <\ 1\,,
\end{equation}
hence it is natural to fit our data at finite $L$ to
a constant leading term, which is the thermodynamic limit, plus a power-law scaling correction:
\begin{equation}\label{eq:space-filling-ansatz}
    1-\qlqz = a + \frac{b}{L^c}\,.
\end{equation}

\begin{figure}
    \centering
    \includegraphics[width=\linewidth]{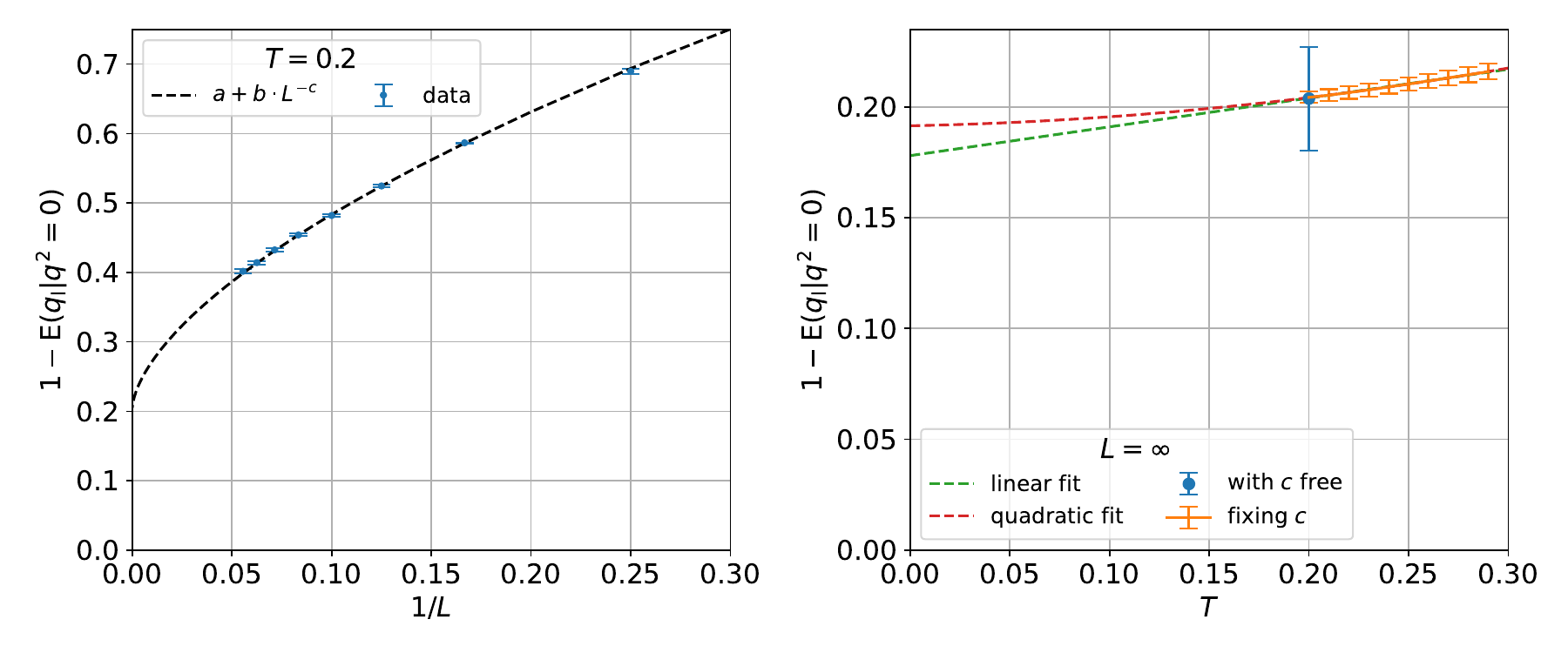}
    \caption{\label{fig:limits}
        {Taking the successive $L \to\infty$ and $T\to0$ limits.}
        \textit{Left:} $\qlqz$ for $T=0.2$  as a function of $1/L$. 
         The dashed line is the fit of these points to Eq.~\eqref{eq:space-filling-ansatz}.
        The resulting parameters, along with the goodness-of-fit metrics, are reported in Table~\ref{tab:scalingfit}. 
        \textit{Right:} $L \to \infty$ values from fits of data at different temperatures to 
        ansatz~\eqref{eq:space-filling-ansatz}. The blue dot represents the result from the fit shown 
        in the left panel, 
        the orange dots are obtained by fitting the interpolated data to the same function but fixing parameter $c$ to the value in Table~\ref{tab:scalingfit}. 
        Green and red dashed lines are the linear and quadratic fits of the orange points, respectively.
    }
\end{figure}

The left panel of Fig.~\ref{fig:limits} shows the result obtained for $T=0.2$, the temperature for which we have data directly from simulations for all 
sizes (aside  from $L=4$ and  $6$ where, as we have already explained, we need to carry out interpolations). 
The blue dots represent the data points, the black dashed curve is the fit to the  ansatz~\eqref{eq:space-filling-ansatz}.
The parameters are reported in Table~\ref{tab:scalingfit}.  As one can notice from the $\chi^2$ per degree of freedom (d.o.f.) and its resulting $p$ value, 
the fit to~\eqref{eq:space-filling-ansatz} has an excellent agreement with the data.

\begin{table}[tb]
\centering
\begin{tabular}{c c c c c c c}
\hline
\hline
          $a$               & $b$               & $c$               & $\chi ^2/\text{d.o.f.}$ & $p$-value               \\ \hline
$0.204 (23)$ & $1.145 (29)$ & $0.612 (47)$ & $4.62/5$      & $0.465$               \\ \hline \hline
\end{tabular}
    \caption{\label{tab:scalingfit}
        Parameters and goodness-of-fit estimators obtained by fitting the data of $T=0.2$ in the left panel of Fig.~\ref{fig:limits} to  ansatz~\eqref{eq:space-filling-ansatz}.
    }
\end{table}

Once we have ascertained that ansatz~\eqref{eq:space-filling-ansatz} fits our data at $T=0.2$, we fix scaling exponent $c$ to the value we got 
for $T=0.2$ ---the one reported in Table~\ref{tab:scalingfit}--- and used the values of $\qlqz$ for each $L$
and for each interpolated $T\in [0.2, 0.3)$ 
to fit to the same function; see Sec.~\ref{subsec:qsq0}. 
The $L\to\infty$ limit, \emph{i.e.}, the fitted parameter $a$, is the thermodynamic limit of $1-\E(q_\text{l}|q^2=0)$. The value of $a$ 
obtained with the first fit, leaving $c$ as a free parameter, is represented in the right panel of Fig.~\ref{fig:limits} as the blue dot, while the values obtained by
fixing $c$ are the orange dots. The new value of $a$ obtained for the data at $T=0.2$ by fixing $c$ is $a=0.2043 (26)$, 
which is perfectly compatible with the value in Table~\ref{tab:scalingfit} but has a smaller error ($\sim 1/9$ of the first one), 
as one would normally expect when there is an additional degree of freedom in the fit. 

From this thermal dependence of $\qlqz$ at $L\to\infty$ we have finally been able to extrapolate the $T\to 0$ limit. 
We did this by fitting again the obtained points to a linear and a quadratic function of $T$, respectively the green and the red dashed 
lines in right panel of Fig.~\ref{fig:limits}. We repeated the operation again fixing $c$ to the upper and lower bounds given by the error 
reported in Table~\ref{tab:scalingfit}. It emerged that the choice of $c$ produces a systematic bias of $\sim 0.02$. 
Thus, our final estimates for the intercepts are $a_0 = 0.1780 (5)[235]$ in the case of the linear fit, and $b_0 = 0.1915 (18)[212]$  for the 
quadratic fit. The first error is statistical, while the one in square brackets is given by the systematic effect from the 
choice of $c$.
Thus our final estimates are 
\begin{align}\label{eq:our_result}
        \text{linear: } \ &\E \left (\davg{q_\text{l}} | q^2=0 \right) \bigg\rvert_{ T\to 0, L\to\infty}=0.8210 (5)[235]\,, \\ 
        \text{quadratic: } \ &\E \left (\davg{q_\text{l}} | q^2=0 \right) \bigg\rvert_{ T\to 0, L\to\infty}= 0.8085 (18)[212]\,.    
\end{align}
After taking the systematic error into account, the two extrapolations to $T=0$ turn out to be mutually compatible, and incompatible with 1, which indicates space-filling surfaces for the relevant excitations.   Sec.~\ref{subsec:otherworks} discusses this finding in relation to previous work.

\subsubsection{The thermodynamic and zero-temperature limits for the energy density}\label{subsubsec:E-limit}

In this section we estimate the ground-state energy in the thermodynamic limit. This quantity has already been computed in the past \cite{pal:96, roma:09}, but the method always relied on 
computing the $L\to\infty$ limit directly from the ground states, hence performing the $T\to 0$ limit first. Again, 
our low-temperature data allows us to take the limits in the opposite order.

To compute the thermodynamic limit, we need to fix $T$ and conduct a scaling analysis.
Clearly, we encounter the same issue we discussed in Sec.~\ref{subsec:qsq0}, that is, data for different sizes is not 
always collected for the same set of temperatures, see Appendix~\ref{sec:appendice_tecnico}. 
Hence, analogously to $\qlqz$, we interpolated our energy measurements. 
To be more precise, let us define
\begin{equation}\label{eq:eL-def}
    e_L(T) = \frac{\davg{\mathcal{H}}}{V}\,.
\end{equation}
We carry out the interpolation in the $0.2\leq T < 0.5$ temperature interval using cubic splines.\footnote{Specifically, we carry out two interpolations, one for $e_L(T_i)$ plus the error bar, the other for $e_L(T_i)$ minus its error bar ($T_i$ are the simulated temperatures). Afterwards, we take $e(T)$ as the average of the two interpolations. Our error estimate is half the difference between the two interpolations.}

\begin{figure}[tb]
    \centering
    \includegraphics[width=0.5\linewidth]{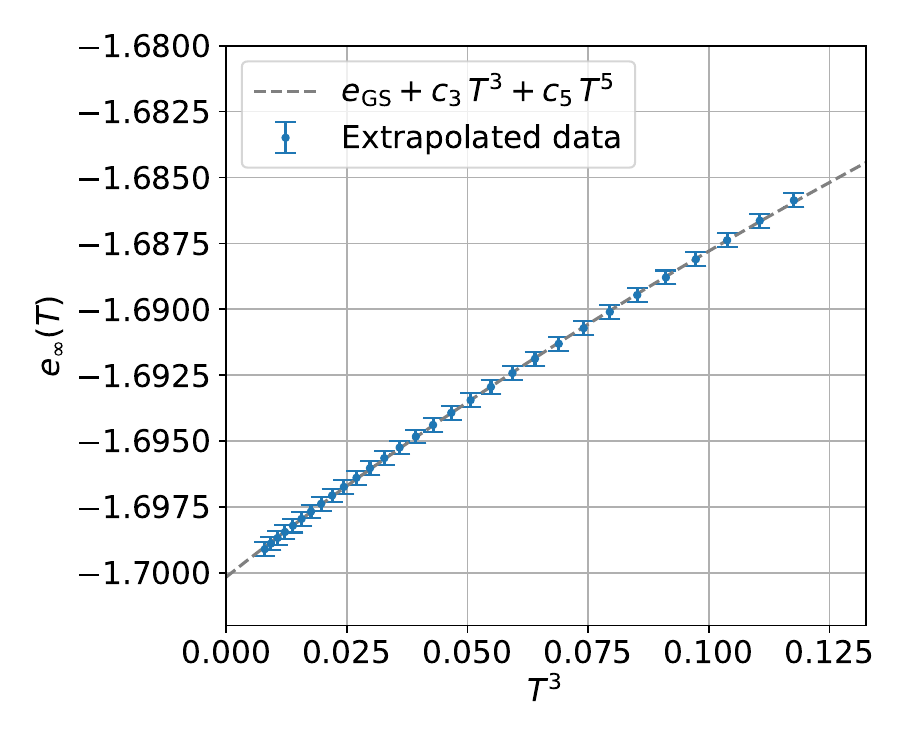}
    \caption{\label{fig:energies-in-T}
    {Energy density as a function of $T$}. 
    Plot of our extrapolations for the thermodynamic limit of the energy density $e_\infty(T)$ obtained at fixed $T$ from the fit to Eq.~\eqref{eq:ene-T-scaling}. The grey dashed line is a fit to Eq.~\eqref{eq:enes-odd-pows}, the corresponding values and accuracy parameters are reported in Table~\ref{tab:energy-gs-fit}.
    }
    
\end{figure}

Next, the interpolated data at fixed $T$  is extrapolated to the thermodynamic limit through an ansatz 
analogous to that of Eq.~\eqref{eq:space-filling-ansatz}:
\begin{equation}\label{eq:ene-T-scaling}
    e_L(T)=e_\infty(T)+\frac{A}{L^B}\,.
\end{equation}
The parameters of the fit are $A=1.70(2)$ and $B=2.78(9)$ for all temperatures.\footnote{Both $A$ and $B$ and their reported errors are the averages over the values obtained in the individual fits, because no noticeable temperature dependence was identified in the fit parameters.} 
Our value for exponent $B$ is consistent with the finding in~\cite{boettcher:24}, 
see Table 1 of that work, 
which is $\omega D=2.745(12)$. On the other hand, both $A$ and $B$ are respectively 3.76 and 3.58 standard deviations from 
the ones obtained in~\cite{roma:09}, see their Table 2, for the ansatz of their Eq.~(12) analogous to 
our ansatz~\eqref{eq:ene-T-scaling}. Our extrapolated values $e_\infty(T)$ are shown in Fig.~\ref{fig:energies-in-T}. 

Our next task is extrapolating our data to $T=0$. Two facts are relevant to that end. First, when the temperature goes to zero $\mathrm{d}e_\infty/\mathrm{d} T\to 0$ as well, as implied by the third law of thermodynamics. Second, because $e_L(T)$ is an odd function of $T$, $e_\infty(T)$ should also be odd. The argument for the symmetry $e_L(T)=-e_L(-T)$ is quite simple. If the lattice is bipartite the partition function is an even function of $T$
(just change variables as $\sigma_i\to -\sigma_i$ for one of the two sublattices). Because the partition function yields $\langle \mathcal{H}\rangle_T$
by taking the derivative of
the logarithm of the partition function with respect to $\beta=1/T$, the average energy is an odd function  of $T$ on a sample-by-sample basis. To extend the argument to a general lattice, one notices that exchanging $T$ by $-T$ amounts to changing the sign of all the couplings $J_{ij}$ in the 
Hamiltonian, Eq.~\eqref{eq:hamiltonian}. Hence, because our coupling distribution is symmetrical under $J \to -J$, the disorder-averaged free energy is an even function of $T$ also for non-bipartite lattices (such as the 
interaction graph for the Sherrington-Kirkpatrick model).

Before presenting our ansatz for the $T\to 0$ extrapolation, there is a final subtle point to discuss, namely that $e_\infty(T)$ is discontinuous at $T=0$:
 \begin{equation}\lim_{T\to0^+} e_\infty(T) \equiv e_\text{GS}=-\lim_{T\to0^-} e_\infty(T)\,.\end{equation}
If we consider instead
\begin{equation}\tilde{e}(T)=e_\infty(T) - \sgn (T) e_\text{GS},\end{equation}
we have a function which is continuous at $T=0$ and odd in $T$. Therefore, making the additional assumption that $\tilde{e}(T)$ can be Taylor expanded at
$T=0$ we finally arrive at our ansatz for the $T\to 0^+$ extrapolation:
\begin{equation}\label{eq:enes-odd-pows}
    e_\infty(T)=e_\text{GS} + c_3\, T^3 + c_5 \, T^5\,,
\end{equation}
where we have neglected contributions of order $\mathcal{O}(T^7)$ and we have set $c_1=0$ due to the third law of thermodynamics. 
Eq.~\eqref{eq:enes-odd-pows} is consistent with the findings for the Sherrington-Kirkpatrick model~\cite{thouless:77,parisi:79b}. We note \emph{en passant} 
that the DM prediction $\tilde{e}(T)=\mathcal{O}(T^2)$~\cite{fisher:88b} implies that $\tilde{e}(T)$ \emph{cannot} be Taylor expanded at $T=0$.

We are finally ready to extrapolate $e_\infty(T)$ to $T=0^+$. The parameters resulting from the fit to Eq.~\eqref{eq:enes-odd-pows} are reported in 
Table~\ref{tab:energy-gs-fit}. The extremely low value of the $\chi^2/\text{d.o.f.}$ parameter is a consequence of data correlation at different temperatures, which was to be expected, since the same samples are simulated in a Parallel Tempering chain. Our extrapolation to $T=0^+$ nicely agrees with previous estimates at exactly $T=0$. Our extrapolation of $e_\text{GS}$ agrees within less than one standard deviation
with the results at zero temperature from Ref.~\cite{pal:96} of $e_\text{GS}=-1.7003(8)$ and with  the estimate from~\cite{roma:09} that uses the same ansatz as we do [their Eq.~(12), same as our Eq.~\eqref{eq:ene-T-scaling}], amounting to $e_\text{GS} = -1.7000(3)$.
The other two ans\"atze for the $L\to\infty$ extrapolation in~\cite{roma:09} are also in agreement with us, namely the extrapolation from
their Eq.~(14) is $e_\text{GS} = -1.7004(2)$ and the extrapolation from their  Eq.~(13) is  $e_\text{GS} = - 1.6997(2)$, which is 1.8 standard deviations from ours.

Finally, a DM-inspired extrapolation to $T=0^+$, namely $e_\infty(T)=e_\text{GS} + c_2 T^2 +c_4 T^4$, nicely interpolates our data ($\chi^2$/d.o.f.=0.086/27, $p$-value=1) but provides too-low an extrapolation, $e_\text{GS}=-1.7010(2)$.

\begin{table}[tb]
\centering
\begin{tabular}{c c c c c} \hline \hline
$e_\text{GS}$ & $c_3$    & $c_5$    & $\chi^2/\text{d.o.f.}$ & $p$-value \\
\hline
$-1.70016(15)$  & $0.148(8)$ & $-0.113(34)$ & $0.625/27$ & $1$ \\
\hline \hline
\end{tabular}
\caption{\label{tab:energy-gs-fit}
    Values and goodness-of-fit parameters obtained by fitting the data in Fig.~\ref{fig:energies-in-T} to the ansatz~\eqref{eq:enes-odd-pows}.
}
\end{table}

\section{Discussion and conclusions}\label{sec:final}

\subsection[$L=18$ is useful]{\boldmath $L=18$ is useful} \label{subsec:useof18}


Although our $L=18$ simulation did not reach equilibrium, it is quite close to attaining it. To frame the discussion, we take a $\log_2$-binning approach to our data. The time series produced by our Monte Carlo simulation is split into logarithmic bins:
\begin{equation}\label{eq:time-bin-def}
\text{bin}(b)=\Big[\frac{t_\text{max}}{2^{b+1}}+1, \frac{t_\text{max}}{2^{b}}\Big]\,.
\end{equation}
Here $t_\text{max}$ is the total number of elementary Monte Carlo steps. 
The second half of the simulation corresponds to bin 0, the second quarter with bin 1, etc.

Table~\ref{tab:sdy-compare-16-18} displays the value of the SDY observable defined in Eq.~\eqref{eq:SDY-def} at $T=T_\text{min} =0.2$ for the $\log_2$ bins of $L=16$ and for the second half of the $L=18$ bin 0  (a problem with the hard-drive system caused the loss of earlier $L=18$ datapoints; that is, for $L=18$ we only have data for the last quarter of the run):
\begin{table}[tb]
\centering
\begin{tabular}{c c c }
\hline \hline
$L$ & bin & $\davg{\SDY}$        \\\hline 
16  & 3   & 0.0279(58) \\
16  & 2   & 0.0194(56) \\
16  & 1   & 0.0073(54) \\
16  & 0   & 0.0033(53) \\
18  & 0 (second half)   & 0.0161(58) \\ \hline \hline
\end{tabular}
\caption{\label{tab:sdy-compare-16-18}
SDY observable defined in Eq.~\eqref{eq:SDY-def} at $T=T_\text{min} =0.2$ for the $L=16$ logarithmic time bins (0: second half of the simulation, 1: second quarter, \ldots)  and for the second half of the $L=18$ bin 0 in our simulations produced with the  PTHR algorithm \cite{chilin:26b}.
}
\end{table}

It is clear from Table~\ref{tab:sdy-compare-16-18} that  (the second half of) bin 
0 for the $L=18$ simulation is somehow intermediate between  bins 1 and 2 of our simulation for $L=16$. Hence, it is natural to ask whether some 
results from this simulation can be used in the analysis. As we argue below, the most obvious candidate for a useful quantity is the conditional expectation $\E(q_\text{l} | q^2)$; see Eq.~\eqref{eq:conditional-expectation}. 

The physical reason underlying this optimistic expectation is overlap equivalence~\cite{contucci:06}: as explained in 
Sec.~\ref{subsec:yes-q2-filter}, when the size of the system approaches 
infinity, the conditional variance $\text{Var}(\ql|q^2)$ vanishes in the thermodynamic limit, 
and the conditional expectation value $\E(\ql|q^2)$ becomes a one-to-one function of $q^2$. This should happen \textit{regardless} the of the equilibration of the data. According to this interpretation, the failure to satisfy  Eq.~\eqref{eq:SDY} would be entirely due to a lack of equilibration of $P(q)$. Indeed,
the positive value of $\davg{S}$ indicates that the $\davg{\ql}$ reached in the simulation is smaller than its equilibrium value. But, recall Eq.~\eqref{eq:sum-rule-1}, $\davg{\ql}$
depends \emph{solely} on  $P(q)$ and on the conditional expectation value $\E(q_\text{l} | q^2)$.

\begin{figure}[t]
  \centering
  \includegraphics[width=\linewidth]{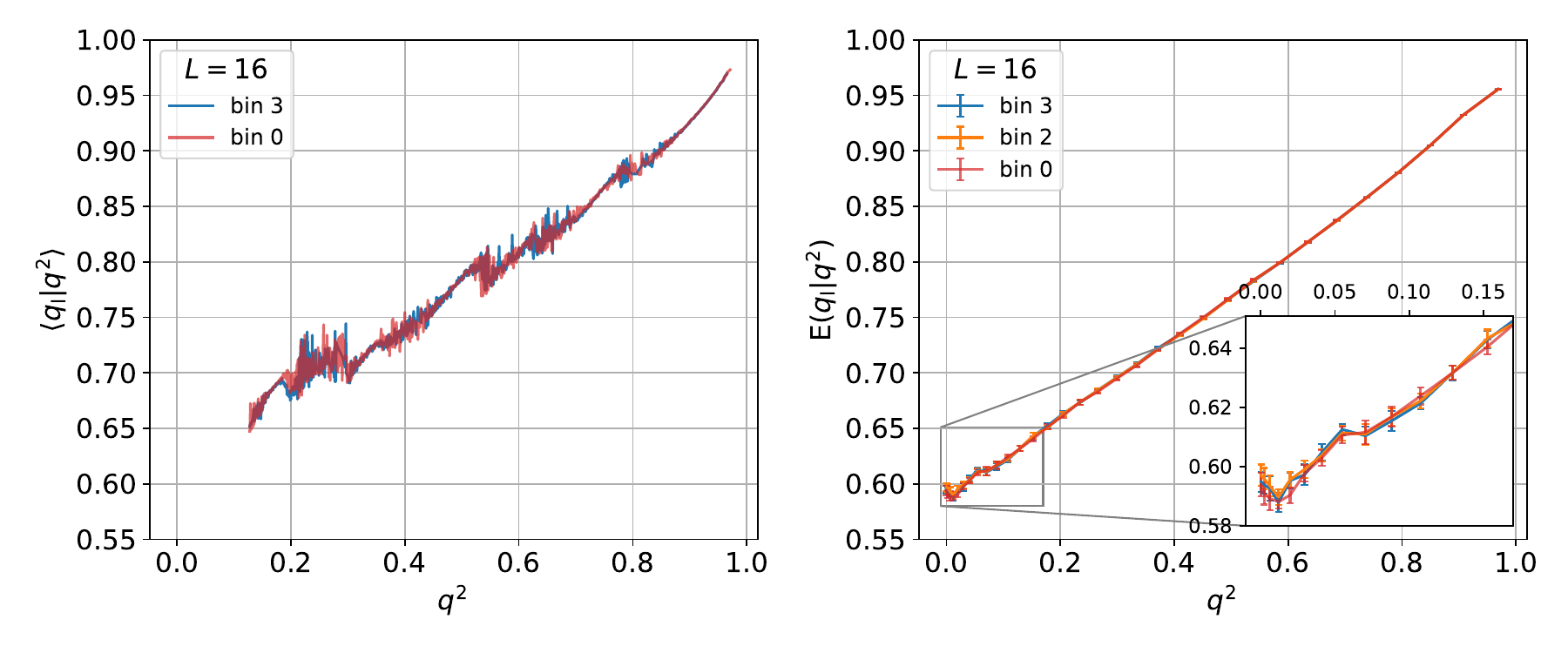}  
\caption{\label{fig:compare_rndtrps-qlqsq}
    {Link overlap $\ql$ conditioned to $q^2$ at different stages of equilibration.}  
    \textit{Left:} Comparison of a single-sample $\tavg{\ql | q^2}=\tavg{\ql \ \delta_{q^2, c^2}}/\tavg{\delta_{q^2, c^2}}$
    between bin 0 (second half of the simulation, equilibrated) and bin 3 (second sixteenth of the simulation, off equilibrium) at $T=0.2$.
    \textit{Right:} Sample-averaged conditional expectation $\E(q_\text{l} | q^2)$, Eq.~\eqref{eq:conditional-expectation}, as a function of $q^2$, as computed for $T=T_\text{min}=0.2$ and $L=16$. See Eq.~\eqref{eq:time-bin-def} for the definition of the $\log_2$-binning and Appendix~\ref{sec:appendice_tecnico} for simulation parameters (only bin 0 is fully equilibrated). The inset is a closeup of the low-$q^2$ area, where fluctuations are strongest.}
\end{figure}

We test our hypothesis in Fig.~\ref{fig:compare_rndtrps-qlqsq}, which compares equilibrated (bin 0 and, perhaps, also bin~1) with non-equilibrated (bins 2 and 3) determinations of $\E(\ql|q^2)$ from our $L=16$ simulations. As the reader will notice, the conditional expectation $\E(\ql|q^2)$ is completely blind to the lack of equilibration. Instead, as Table~\ref{tab:quartiles} shows, lack of equilibration is quite visible for 
both $\davg{\ql}$ and for $P(q)$, which we investigate through $\davg{q^2}$ and through the cumulative distribution $\mathcal{F}(q_{\text{max}}^2)$ of Eq.~\eqref{eq:prob-q2-def}. 

On the view of these results, we have decided to keep our estimates of  $\E(q_\text{l} | q^2)$  from  the $L=18$ simulation  in the analysis.
\begin{table}[t]
\centering
\begin{tabular}{c . . . . }
\hline
\hline
Quantity & \multicolumn{1}{c}{Bin 3} & \multicolumn{1}{c}{Bin 2} & \multicolumn{1}{c}{Bin 1} 
&\multicolumn{1}{c}{Bin 0}        \\ \hline
$\davg{q^2}$         & 0.783 (3)   & 0.787(3)   & 0.793(3)   & 0.795(3)   \\ 
$\davg{q_\text{l}}$         & 0.8812(12) & 0.8829(11) &0.8853(11) & 0.8861(11) \\ 
$\mathcal{Q}_1$ & 0.78(5) & 0.783(4)     & 0.790(4) & 0.794(4)     \\ 
$\mathcal{Q}_2$    & 0.8765(11) & 0.8783(10) & 0.8799(10) & 0.8809(9)     \\ 
$\mathcal{Q}_3$ & 0.9136(5) & 0.9142(5)     & 0.9150(5) & 0.9153(5)     \\ \hline
\hline
\end{tabular}
\caption{\label{tab:quartiles}
   {Probability distribution $P(q)$ and $\davg{\ql}$ at different stages of thermalization.} We present the disorder averages of $q^2$ and $q_\text{l}$ [Eqs.~\eqref{eq:tot_ovlp} and~\eqref{eq:link_ovlp}], along with the first and third quartiles $\mathcal{Q}_1, \mathcal{Q}_3$ and the median $\mathcal{Q}_2$ of the cumulative distribution function $\mathcal{F}(q_{\text{max}}^2)$, Eq.~\eqref{eq:prob-q2-def}, as computed from the different time bins of our PTHR simulation for $L=16$ at $T=T_\text{min}=0.2$. Time bins are defined in Eq.~\eqref{eq:time-bin-def}. Errors are computed using the bootstrap method.
    }
\end{table}


\subsection{Discussion: comparison with previous work}\label{subsec:otherworks}

We start by making the crucial point that our computation of $\qlqz$ at $T=0.2$ can be regarded as a $T=0$ determination, particularly for large sizes (Sec.~\ref{subsec:TfiniteT0}).  

Afterwards, we compare our results to those obtained in previous works that used  different methods. The first natural classification separates works that do not ensure that the volume and the surface under consideration actually belong to the
same excitation ---Refs.~\cite{marinari:00d,palassini:00,shen:24}, see Sec.~\ref{subsec:no-q2-filter}--- from works that ensure that the same 
excitation is under  study ---Ref.~\cite{marinari:01,marinari:02, palassini:03}, see Sec.~\ref{subsec:yes-q2-filter}. This does not imply that the results 
of  Refs.~\cite{marinari:00d,palassini:00,shen:24}  are wrong, it just means that their interpretation is less straightforward.

\subsubsection[Is it reasonable that $\qlqz$ is so similar for $T=0$ and $T=0.2$?]{Is it \boldmath reasonable that $\qlqz$ is so similar for $T=0$ and $T=0.2$?}\label{subsec:TfiniteT0}

\begin{table}[tb]
\centering
\begin{tabular}{r . . . .}
\hline
\hline
\multirow{2}{*}{$L$} & \multicolumn{1}{c}{$1-\qlqz$} & \multicolumn{1}{c}{$1-\qlqz$}            & \multicolumn{1}{c}{\multirow{2}{*}{Difference}}& \multicolumn{1}{c}{\multirow{2}{*}{Difference/error}}            \\ 
  & \multicolumn{1}{c}{$T=0.2$}          & \multicolumn{1}{c}{$T=0$~\cite{marinari:01}}              &       &                     \\  \hline
4                       & 0.6894(37) & 0.7224(9) &-0.0330                & -8.724        \\
6                       & 0.5864(7) & 0.6006(5) & -0.0142      & -17.16                \\
8                       & 0.5227(48)& 0.5347(22) & -0.0120                & -2.267        \\
10                      &0.4814(15)& 0.4875(23) & -0.0061   & -2.211             \\
12                      & 0.4530(22)       & 0.4593(22)  &-0.0063               & -2.016\\
14                      & 0.4318(20) & 0.4313(42)  &0.0005           & 0.1084\\
16                      & 0.4135(24)               & & & \\ 
18                      & 0.4018(31)                & & & \\ \hline
\hline
\end{tabular}
\caption{\label{tab:mp_comparison}
    Comparison between the values computed herein at $T=0.2$, and the ones for $T=0$ obtained in Ref. ~\cite{marinari:01} though a bulk perturbation to the ground state. The differences between the $T=0.2$ and the $T=0$ computations are given in absolute value and in units of the statistical error.}
\end{table}

Our values of $\qlqz$ at $T=0.2$ are compared with the $T=0$ results of~\cite{marinari:01} in Table~\ref{tab:mp_comparison}.
From the table it is apparent that the distance between the estimates at low but finite temperature and $T=0$ 
is small, and shrinks when $L$ grows. We present here an interpretation of this surprising similarity that is based on overlap equivalence~\cite{contucci:06}.

Overlap equivalence depends on two main hypothesis: (i) the conditional variance $\text{Var}(\ql|q^2)$ vanishes in the thermodynamic limit, 
and (ii) the conditional expectation value $\E(\ql|q^2)$ becomes a one-to-one function of $q^2$ when $L\to\infty$. 

As far as we know, hypothesis (i) is undisputed. Indeed, numerical investigations have reported 
$\text{Var}(\ql|q^2)\propto L^{-D/2}$~\cite{contucci:06,janus:10}. From the opposite side in this controversy, TNT supporters claim that the 
standard variance for $\ql$ vanishes in the thermodynamic limit~\cite{katzgraber:01}, which, combined with the sum 
rule of Eq.~\eqref{eq:sum-rule-2}, implies that the conditional variance also  goes to zero.

The second hypothesis of overlap equivalence is, instead, at the root of the controversy. Although $\E(\ql|q^2)$ obtained at finite $L$ is 
indeed a one-to-one function of $q^2$, see Figs.~\ref{fig:pq_qlqsq} and \ref{fig:fits_and_interpols}, a TNT proponent would object that the 
slope in these figures decreases as $L$ grows. However,  see Table~\ref{tab:c_comparison}, in $D=3$ the slope decreases with $L$ much more slowly 
than $\sqrt{\text{Var}(\ql|q^2)}\propto L^{-D/4}$. Hence, we think it is uncontroversial to state that, at least for finite $L$, the \ql\
computed on a pair of independent configurations in thermal equilibrium can be approximated as
\begin{equation}\label{eq:overlap-equivalence}
\ql\simeq \qlqz + a_2 q^2+\mathcal{O}(q^4)\,,
\end{equation}
for small $q^2$, where the $\simeq$ symbol accounts for the non-vanishing (but small) conditional variance. The final observation to conclude the argument 
comes from Fig.~\ref{fig:fits_and_interpols}--right, where the reader can notice the tiny, almost negligible, temperature dependence of $\qlqz$. 

At this point, the results in Table~\ref{tab:mp_comparison} look natural: Since the bulk perturbation employed in Ref.~\cite{marinari:01} was designed to 
generate an excess energy over the ground state of $\mathcal{O}(1)$, both the ground state  and the perturbed ground state are typical configurations for 
the Boltzmann distribution at low-enough $T$, which means that Eq.~\eqref{eq:overlap-equivalence} should be valid if the   value of $q^2$ computed from 
the two  ground states turns out to be small enough. Or, the other way around (and perhaps more precisely), Table~\ref{tab:mp_comparison} can be seen as 
evidence that the bulk-perturbation method, as applied to the lattice sizes in the table, generates configurations that are typical for the Boltzmann 
distribution at low-enough temperatures.

\subsubsection[Works that do not filter with $q^2$]{Works that \boldmath do not filter with $q^2$}\label{subsec:no-q2-filter}

\begin{table}[t]
\centering
\begin{tabular}{c c c c c}
\hline \hline
$L_\text{min}$ & $L_\text{max}$ & $c_\text{eff}      $& error     & $\chi^2/\text{d.o.f.}$ \\ \hline
4    & 6    & 0.399 & 0.013  &           \\
6    & 8    & 0.400 & 0.032  &           \\
8    & 10   & 0.369 & 0.044  &           \\
10   & 12   & 0.334 & 0.032  &           \\
12   & 14   & 0.311 & 0.044  &           \\
14   & 16   & 0.325 & 0.056  &           \\
16   & 18   & 0.244 & 0.082  &           \\\hline \hline
4    & 12   & 0.383 & 0.005 & 4.11/3 \\
6    & 14   & 0.372 & 0.004 & 9.85/3 \\
8    & 16   & 0.329 & 0.011 & 1.10/3 \\
10   & 18   & 0.317 & 0.010 & 1.16/3 \\ \hline \hline
\end{tabular}
\caption{
    \label{tab:c_comparison}
    Size-dependent, effective estimates of the fractal dimension of the excitations from Eq.~\eqref{eq:trivial-ansatz}, where exponent $c$ plays the role of $D-\df$. In the top rows,
    $c_\text{eff}$ is computed from the two-system-size estimator in Eq.~\eqref{eq:ceff} and, therefore, no
    $\chi^2$ value is reported since no fit was performed. 
    In the bottom rows, $D-\df$ is estimated from fits of our data for $T=0.2$ to 
    the trivial-$\ql$ ansatz~\eqref{eq:trivial-ansatz}, restricted to intervals $L\in [L_\text{min},L_\text{max}]$. 
}
\end{table}

We start with the results of Ref.~\cite{marinari:00d}. This work studies the ground states of samples up to $L=14$ before and after changing 
the boundary conditions from periodic to antiperiodic. As stated above, these authors studied the behavior of the link 
overlap, Eq.~\eqref{eq:link_ovlp}, without conditioning to  low values of the spin overlap $q$, thus including  small-size excitations in 
their computation. Their final extrapolation to infinite system size is $\davg{\ql}=0.765(15)$, which is lower than, although almost 
compatible with, our own estimate in  Eq.~\eqref{eq:our_result}. As explained in the Introduction, we should stress that varying the boundary 
conditions induces excitations with an energy variation which is \textit{not} guaranteed to be of $\mathcal{O}(1)$.  

Ref.~\cite{palassini:00} considered another way of inducing ground-state perturbations that were certain to raise the energy by an $\mathcal{O}(1)$ amount, finding in this way $D-\df=0.42(2)$.  We recall that we have shown in Sec.~\ref{subsec:TfiniteT0} that our  $\qlqz$ computed at $T=0.2$ can be regarded as representative of the $T\to 0$ limit also for finite sizes. Therefore,  we can use this data to test the TNT theory, where a power law describes how $\qlqz$ approaches $1$ when $L\to\infty$:
\begin{equation}\label{eq:trivial-ansatz}
    1-\qlqz = \frac{b}{L^c} \quad \text{(trivial-$\ql$ ansatz)}.
\end{equation}
The fit to the above ansatz is untenable: $\chi^2/\text{d.o.f.}=31.8/6$ ($p=1.78 \times 10^{-5}$). Even though we can rule out the simple power law form of Eq.~\eqref{eq:trivial-ansatz}, we still cannot exclude that it provides the leading term for  a more complicated expression containing some form of corrections. In order to test this, we fitted this data
to a function of the form $d/L^e(1+b/L^c)$. The fit is fair ($\chi^2/\text{d.o.f.}=4.74/4$) but the value for the leading-term exponent is $e = 0.02(63)$. This result, very close to 0, suggests again the non-triviality of $\ql$ for the EA3D model. The error
is, however, too large completely to rule out a power-law scaling of
the leading term. A more refined analysis is needed.

We start with an effective estimate of exponent $c$ in Eq.~\eqref{eq:trivial-ansatz}, which plays the 
role of $D-\df$, obtained from just 
two system sizes $L_\text{max}$ and $L_\text{min}$:
\begin{equation}\label{eq:ceff}
    c_\text{eff}(L_\text{max},L_\text{min})=-\frac{\log \E (\ql | q^2=0, L_\text{max})-\log\E (\ql|q^2=0, L_\text{min})}{\log L_\text{max}-\log L_\text{min}}\,.
\end{equation}
When $L_\text{max}\leq 8$ we find results for $c_\text{eff}$ in excellent agreement with the 
results of Refs.~\cite{palassini:00}; see Table~\ref{tab:c_comparison}. However, when the considered sizes increase, the estimate of $c_\text{eff}$ diminishes. 
Unfortunately, the error estimate increases as well. This is why we have included in Table~\ref{tab:c_comparison} a second estimate of
$c_\text{eff}$ from a fit to Eq.~\eqref{eq:trivial-ansatz} limited to five consecutive system sizes. As the window $[L_\text{min},L_\text{max}]$ slides to 
larger sizes, smaller values of $c_\text{eff}$ are found. 

Establishing the equality of two numbers ---$D$ and $\df$ in our case--- is beyond the capabilities of any numerical or experimental investigation, the best one can do is set bounds. The $L_\text{max}=8$ investigations put this bound at $D-\df\leq 0.42$~\cite{palassini:00}, while
the present investigation with $L_\text{max}=18$ has taken the bound to $D-\df\leq 0.32$. On the other hand, see Sec.~\ref{subsec:limits}, our data is fully consistent with the  $D=\df$ hypothesis.

Finally, Ref.~\cite{shen:24} considers a different way of producing excitations with $\mathcal O(1)$ excess energy that they study by computing  ground states. These authors strongly perturb a single 
link of the system (instead, Ref.~\cite{palassini:00} weakly perturbs \emph{all} links in the system). They study in this way system sizes up to 
$L=12$, finding a barely normalizable power-law  distribution for the surfaces of the excitations. Should this behavior persist for larger 
system sizes, it would imply that the median excitation produced by this method ---or indeed the 99th-percentile excitation--- has a surface that 
remains of order $1$ as $L$ grows. Only the effective cutoff diverges with $L$ (the effective cutoff represents the largest surfaces that appear when 
perturbing a system of size $L$; these are rare events that occur with a probability that vanishes as $L\to\infty$). This is in stark contrast with the excitations relevant for $T>0$, for which the scaling 
of the 
conditional
variance $\text{Var}(\ql | q^2)\propto 1/L^{D/2}$~\cite{contucci:06,janus:10} indicates that  large statistical fluctuations ---let alone rare events--- do not play a significant role. Under these circumstances, the relevance ---if any--- of the excitations found through single-link perturbation  needs clarification.

\subsubsection[Works that filter with $q^2$]{Works that \boldmath filter with $q^2$}\label{subsec:yes-q2-filter}

An interesting follow-up to~\cite{marinari:00d} is in Ref.~\cite{marinari:01}, where the bulk perturbation of~\cite{palassini:00} is employed. Ref.~\cite{marinari:01} improves over~\cite{palassini:00} by simultaneously analyzing the site, Eq.~\eqref{eq:tot_ovlp}, and link overlap, Eq.~\eqref{eq:link_ovlp}, for the original and the perturbed ground states
of samples up to $L=14$. The step forward, as compared to~\cite{marinari:00d}, is that the excess energy of the excitation is  of $\mathcal{O}(1)$ by construction. By considering small values of $q^2$, these excitations 
are maximal in volume by definition. 
Moreover, they focus the study by conditioning to $q^2<0.3$. The main issue with Ref.~\cite{marinari:01} is that, using different methods, they extrapolate to four different values of $\qlqz$, which are 
mutually incompatible and are all lower than our own extrapolation.
It is interesting to note though that their fit of the $1-\qlqz$ data
to ansatz~\eqref{eq:space-filling-ansatz},
which is the method most similar to ours, gives $1-\qlqz=0.24 (1)$, which is 
within 2 error bars from our fitted value of $a$ for $T=0.2$ reported in Table~\ref{tab:scalingfit} [our extrapolations for $1-\qlqz$ are 
essentially $T$-independent, see Fig.~\ref{fig:limits}-right]. This coincidence at the $2\sigma$ level has been discussed in Sec.~\ref{subsec:TfiniteT0}.

It is interesting to note Ref.~\cite{palassini:03}, which follows the approach of ~\cite{palassini:00} but with some important changes:
First, they condition their measurements to the cases in which the overlap between the original and the 
perturbed ground state is $|q|\leq 0.4$. Secondly,
they compute the ground states in an exact way, up to $L=10$, but in order to do so they employ free boundary conditions,
which enhance finite-size effects.
Their results are compatible with the space-filling hypothesis, although alternative non-space-filling fits are also tenable.

Finally, we  compare with the  off-equilibrium study  reported in Ref.~\cite{marinari:02}. In the dynamic approach, one should
take three limits with a very precise ordering: (i) the thermodynamic limit, (ii) the infinite-time limit, and (iii) the $T=0$ limit.
The authors of Ref.~\cite{marinari:02} took the limits (i) and (ii). We shall briefly elaborate on their data, in order to take the final 
$T\to 0$ limit. These authors studied systems with size $L=64$ down to temperature $T=0.3$. Just as in our case, they do not 
prescribe a particular type of excitations but let them arise naturally from the dynamics. They are confined to the $q=0$ sector from the 
outset, since the correlation length is smaller than the lattice size. At $T=0.3$, we may compare directly our equilibrium estimate for the 
$L\to\infty$ limit $\qlqz=0.78(2)$ with the infinite-time extrapolation in their Fig.~2, $\qlqz=0.66(4)$~\cite{marinari:02}. The discrepancy is 
large, 2.7 standard deviations, which probably indicates that the infinite-time extrapolation of Ref.~\cite{marinari:02} needs to be 
revisited. As for the extrapolation to $T=0$, we take the data with $T\in[0.3, 0.6]$ from  their Fig. 2 and get, via a linear fit, an estimate  $\qlqz=0.725(7)$, significantly smaller than our own extrapolation to $T=0$ in Eq.~\eqref{eq:our_result}. However, if we restrict ourselves to their data in $T\in[0.3, 0.5]$ the extrapolation  grows to a larger value $\qlqz=0.737(8)$. Hence, we find it plausible that this small systematic discrepancy will be resolved through simulations of larger systems to longer times.

\subsection{Conclusions}

We have investigated the nature of the spin-glass phase by studying the surface-to-volume ratio of system-wide excitations through massive numerical simulations.

Numerical validations of the different theories for the low-temperature spin-glass phase in three spatial dimensions have been queried from two main angles. On the one hand, if one works at finite $T$ it is necessary to find 
smooth extrapolations to $T=0$, but the temperatures at which thermal equilibrium can be reached have been too high. On the other hand, the strategy of going directly to $T=0$ by finding ground states and 
then studying what happens when the system sizes increases is more straightforward (and arguably easier numerically), but is plagued by the problem of 
arbitrariness: One needs to introduce \emph{reasonable perturbations} of the ground state that somehow produce \emph{typical low-energy excitations}, which is 
a completely uncontrolled approximation. We have addressed both problems in this work.

By appropriately modifying Houdayer's cluster method, we have been able to equilibrate lattices of size $L=16$ down to temperatures $T_\text{min}
=0.2\approx 0.21 T_\text{c}$ (and lattices of size $L=18$ were almost equilibrated). The previous  record for this $T_\text{min}$ was 
$L=10$~\cite{katzgraber:07}, although these authors were interested in a different problem. We have shown this $T_\text{min}=0.2$ is low enough to 
extrapolate the energy down to $T=0$. Along the process, we have shown that only the RSB-inspired extrapolation produces the correct value of the ground-state energy. Also, the surface-to-volume ratio of the excitations ---as estimated through the conditional expectation of the link overlap--- is numerically 
coincident at $T=0$ and $T=0.2$, provided the system sizes are large enough.

As for the second problem, we have found in this work ample numerical evidence for the overlap-equivalence property~\cite{contucci:06}, which states that ---for low-lying excited states---  the link overlap is a unique function of the size of the excitation, no matter how the low-lying excited state is found. Overlap 
equivalence provides an \emph{a posteriori} justification of approaches based on perturbing the ground state. The main problem with the ground-state approach, however, is 
that system-wide excitations of low energy become rare at low temperature, as expressed by the PaT scaling that we have studied. Therefore, when the ground 
state is perturbed, system-wide excitations appear as rare events of decreasing probability as the system size increases.

Finally, our simulations are consistent with the equality between the fractal dimension of the excitation surface and the spatial dimension, that is, $\df=D$. While the strict equality of two real numbers is impossible to prove from a numerical calculation, it remains the most economical description of our results.

\section*{Acknowledgements}


\paragraph{Funding information}
This work was partially supported by the Ministerio de Ciencia, Innovación y Universidades (Spain) and by the European Regional Development Fund through grants no. 
 PID2022-136374NB-C21, PID2024-156352NB-I00, PID2024-158623NB-C22 and RED2022-134244-T (MCIU/AEI/10.13039/501100011033/FEDER, UE); by Junta de Ex\-tre\-ma\-dura (Spain) through grant no.~GR24022; and by funding from the 2021 first FIS (Fondo Italiano per la Scienza) funding scheme (FIS783 - SMaC - Statistical Mechanics and Complexity) from
Italian MUR (Ministry of University and Research). We acknowledge the use of the CESAR computational resources at the BIFI Institute (University of Zaragoza) and those of the Instituto de Computacion Cient\'{\i}fica Avanzada de Extremadura (ICCAEx).

\section*{Data availability}

The data contained in the figures in this paper, as well as the Jupyter notebook that generates them, can be downloaded from \href{https://github.com/JeanShc/lowe-excitations-ea3d}{https://github.com/JeanShc/lowe-excitations-ea3d}.

\begin{appendix}
\numberwithin{equation}{section}
\section{Technical details}\label{sec:appendice_tecnico}

This section contains the parameters of the simulations we performed and  of the fits in Fig.~\ref{fig:fits_and_interpols}.

\subsection{Parameters of the simulations}
For $L \leq 12$ we used the regular Parallel Tempering algorithm, while for $L=14, 16, 18$ we used the enhanced procedure we call PTHR, extensively explained in~\cite{chilin:26b}. In the latter case, we used a regular Metropolis algorithm as the local update for the $L=14$ simulations  and for bins 2 and 3 of $L=16$; for bins 0 and 0 of the $L=16$ and the $L=18$ simulations we used Microcanonical Simulated Annealing (MicSA)~\cite{bernaschi:26}; see Eq.~\eqref{eq:time-bin-def} for the definition of time bins.

Table \ref{tab:PTparameters} reports the simulation parameters and the algorithm used for each of the displayed sizes. 
The $N_T$ Parallel Tempering temperatures are uniformly spaced between $T_\text{min}$ and $T_\text{max}$ (both included).
Data is usually averaged for bin 0 of the simulation, aside from $L=18$, where due to technical issues we only have data for the second half of bin 0. 

\begin{table}[h]
\centering
\begin{tabular}{lllllll}\hline\hline
$L$ \ \ & $T_\text{min}$ \ \ & $T_\text{max}$  \ \ & $N_T$ \ \ & $N_\text{samples}$\ \ & $t_\text{max}$         \ \ & Algorithm  \\ \hline
4   & 0.05      & 1.5       & 20        & 25000      & $10^6$            & PT         \\
6   & 0.05      & 1.5       & 20        & 28188      & $2 \times 10^6$    & PT         \\
8   & 0.2       & 1.5       & 13        & 5000       & $2 \times 10^6$    & PT         \\
10  & 0.2       & 2         & 40        & 1536       & $4 \times 10^6$    & PT         \\
12  & 0.2       & 2         & 40        & 1938       & $4 \times 10^6$    & PT         \\
14  & 0.2       & 2         & 48        & 2000       & $6.4 \times 10^6$  & PTHR       \\
16  & 0.2       & 2         & 60        & 2081       & $3.84 \times 10^7$ & PTHR/PTHR* \\
18  & 0.2       & 2         & 72        & 2028       & $3.48 \times 10^7$ & PTHR*      \\ 
\hline\hline
\end{tabular}
\caption{ \label{tab:PTparameters}
    Parameters used for the simulations. PT = Parallel Tempering, PTHR = Parallel Tempering with Houdayer moves and entropy Reservoir, using Metropolis for local updates, 
    PTHR* = Parallel Tempering with Houdayer moves and entropy Reservoir, using MicSA for local updates. In the $L=16$ case, 
    the simulation used PTHR for bins 2 and 3 and then switched to PTHR*.}
\end{table}

\subsection{Technical details on the fits}\label{subsec:appendice_tecnico_q2_to_zero}

In the procedure of extrapolating the value of $\qlqz$ explained in Sec.~\ref{subsec:qsq0}, 
we had to adjust the data window to perform the fit in order to maximize the accuracy. Table~\ref{tab:fit-intervals} reports the used parameters.

\begin{table}[h]
\centering
\begin{tabular}{lllllllll}\hline\hline
L                                             & 4   & 6   & 8    & 10                          & 12   & 14  & 16                        & 18                        \\ \hline
$q^2_\text{min, lin}$                         & 0   & 0   & 0.05 & 0                           & 0    & 0   & 0                         & 0                         \\
$q^2_\text{max, lin}$                         & 0.6 & 0.5 & 0.6  & 0.65                        & 0.7  & 0.7 & 0.7                       & 0.7                       \\
$q^2_\text{min, quad}$ & 0   & 0   & 0    &  0    & 0    & 0   & 0                         & 0                         \\ 
 $q^2_\text{max, quad}$ & 0.7 & 0.5 & 0.8  & 0.65 & 0.75 & 0.7 & 0.75 & 0.70\\ \hline\hline
\end{tabular}
\caption{\label{tab:fit-intervals}
    Intervals used to fit the data as shown in Fig.~\ref{fig:fits_and_interpols}.
}
\end{table}
\end{appendix}



\end{document}